\documentclass[aps,prb]{revtex4}
\usepackage{graphicx,amsbsy,bm}

\begin{document}

\title{Surface and volume plasmons in metallic nanospheres in semiclassical RPA-type approach; near-field coupling of surface plasmons with
semiconductor substrate}
 \author{J. Jacak$^1$, J. Krasnyj$^{1,2}$, W. Jacak$^1$, R. Gonczarek$^1$, A. Chepok$^2$, L. Jacak$^1$}

\affiliation{$^1$Institute of Physics, Wroc{\l}aw University of Technology,
Wyb. Wyspia\'{n}skiego 27, 50-370, Wroc{\l}aw, Poland;\\
$^2$Group of Theor. Physics, International  University, Fontanskaya Doroga 33, Odessa, Ukraine}

\begin{abstract}

 The random-phase-approximation semiclassical scheme for description of plasmon excitations in large metallic nanospheres, with radius range 10--60 nm,
   is formulated in an all-analytical version.
 The spectrum of plasmons  is determined including both  surface and volume type
 excitations and  their mutual connections.  The various channels for damping of surface plasmons are evaluated and the relevant resonance shifts are
 compared with the experimental data for metallic nanoparticles of different size located in dielectric medium or on the semiconductor substrate.
 The strong enhancement of  energy transfer from  the surface plasmon oscillations to the substrate semiconductor is explained  in the regime of a
 near-field coupling in agreement with  recent experimental observations for metallically nanomodified  photo-diode systems.

\end{abstract}
\vspace{1mm}
\vspace{1mm}

\maketitle

\section{Introduction}

Rapid  progress in plasmonics\cite{plasmons} (taking advantage of peculiar properties of plasmon-polaritons\cite{maradudin}
 in  nano-structured metallic interfaces) and plasmonic applications  in photonics and microelectronics\cite{zastos},
focused attention on metallically modified systems in  nano scale and on collective excitations of metallic plasma in a confined
geometry. Of particular interest are
recently reported experimental data on giant enhancement of photoluminescence and absorption of light by semiconductor surface (of photo-diode) covered
with metallic (gold, silver, or copper) nanospheres, with sphere radius of order of several to several tens of nanometers\cite{wzmocn1,wzmocn2,wzr1,wzr2,wzr3,wzr4}.

These phenomena are considered as  perspective for  enhancement of efficiency
of solar cells by application of special  metallic nanoparticle coverings of photo-active layer\cite{wzmocn1,wzmocn2}.
Metallic nanospheres (or  nanoparticles of other shape) can act  as light converters, collecting energy of incident photons in surface plasmon
oscillations.
This energy can be  next transferred  to semiconductor substrate in a more efficient manner in comparison to the direct
photo-effect. The experimental observations\cite{wzmocn1,wzmocn2,wzr1,wzr2,wzr3,wzr4} suggest, that  the short range coupling between plasmons in nanospheres and
electrons in the semiconductor substrate  allows for significant growth of selective light energy transformation into a photo-current in diode system. This
phenomenon is not described in detail as yet, moreover  some competitive mechanisms apparently contribute, manifesting themselves in strong
sensitivity of the effect  to  size and shape of metallic nanocomponents, type of the material and dielectric coverings of nanoparticles\cite{wzr4,wzr5}.
It is probably connected
with particularities of the short range dipole-type (similar to the F{\"o}rster interaction)\cite{wzr1} coupling of
 sphere surface electrical oscillations with semiconductor substrate.
Nevertheless, one can argue
generally
that due to nanoscale of the metallic components the momentum is not conserved, which leads to the allowance of all indirect optical
interband transitions in semiconductor layer, resulting in enhancement of a  photo-current in comparison to the ordinary photo-effect when only direct interband
transitions were admitted. Though the effect is selectively ranged to a
 vicinity of resonant
plasmon frequency, the gain in the total efficiency  would be enlarged by possible dispersing of  dimension and shape (or by a dielectric coating)
of metallic nanoparticles,  widening the resonant spectrum. Any efforts towards improvement of the efficiency rate
of photo-cells via not complicated metallic modifications of their surfaces are of particular significance, as the main drawback on the way for a wider
application of these renewable electricity sources is their relatively small efficiency (reaching not more than 10--20\%,
 depending on a material).

 Since  in  the
metallically modified photo-cell structures the surface plasmons play a central role, the  recognition of these excitations
in nanoparticles is important. The surface plasmons have been originally considered by Mie\cite{mie}, who provided a classical
description of oscillations of electrical charge on the surface of the metallic sphere within the electron-gas model.
The dipole-type Mie surface oscillations of the  electron plasma
are not dependent of the sphere radius, in contradiction to experimental observations both
for small metallic clusters and for larger spheres.
The plasmon effects are of  particular significance in noble metals (gold, silver, and also copper) due to strong visible-light
plasmon resonances in these materials. Even though  a bulk luminescence rate in  metals is small ($\sim 10^{-10}$),
 the rough surface structured in the nano-scale exhibits
luminescence enhancement by several orders of magnitude. A rough metal surface can be treated as a collection
of randomly oriented nano-hemi-spheroids, with distinct spectra of surface plasmons,  increasing
the luminescence rate (e.g., by 6 orders in gold\cite{6orders})---in particular it leads
 to an easily observable 'multi-colored lightning rode effect'\cite{rodeffect} on the surface of broken polycrystalline metallic sample.
Note also that triangle-branched recently synthesized gold colloidal nanoparticles\cite{hao} are blue  in contrast
to spherical ones being  red, which is  caused by the shape-induced shift of surface plasmon resonance\cite{burda}.

In order to describe all these phenomena the plasmon properties beyond the Mie-type picture  need to be recognized.
Plasmon excitations in metallic clusters were the subject of wide analyses\cite{brack,kresin,ekardt,ekardt1,weick,weick1,serra,rubio,gerch}
    within various attitudes including quantum effects.
    The quantum effects
in small metallic clusters were analyzed  using numerical methods
of  calculus 'ab initio', including shell-model and Kohn-Sham 'local density approximation' [LDA], similar as applied in chemistry for large molecule calculations,
limited however to few hundreds of electrons\cite{brack,ekardt,ekardt1,weick,serra}. Also  variational methods for energy density and the random-phase-approximation
(RPA) numerical summations were applied (e.g., for clusters of Na with
radius $\sim 1$nm)\cite{brack1} and various semiclassical expansion methods\cite{kresin,weick,weick1}.
 Emerging of the Mie response from the more general behavior was presented, in particular,  decoupling of
 surface---translational and volume---compressional modes in small clusters was demonstrated\cite{brack,ekardt,brack1}.
In description of metallic clusters commonly the 'jellium' model
was adopted, allowing for adiabatic approach to background ion system. In jellium model all the kinetics concerns
electron liquid screened by static and uniform positive background of ions\cite{brack,ekart,ksi,kre}.

Problem of plasmon oscillations in metallic clusters
was analyzed also including analytical formulation employing  Thomas-Fermi-type approach, e.g., by Kresin\cite{kresin}
or more recently within semiclassical expansion and separation of mass center and relative electron dynamics\cite{weick},
without, however, inclusion of  damping  of surface plasmon oscillations due to radiation losses and addressed rather to small systems
(up to ca 200 atoms, i.e.,  at most 2.5 nm for radius\cite{weick}). The similar range of sphere size was assumed in
more microscopic approach, time-dependent local density approximation (TDLDA), widely applied to metallic
 cluster analyses (e.g., Refs  \onlinecite{brack,ekardt,weick}), where the scheme above Thomas-Fermi
  approximation was developed, including spill-out effect of electron cloud beyond positive jellium,
  screening effects and decay of surface plasmons into particle-hole pairs
  (Landau damping) \cite{ekardt1,weick1}. TDLDA type of approach, though also basing on
  jellium model, in a much better way allows for experimental data  elucidation, especially  for  small and ultra-small clusters,
   with radius up to 1--2 nm, mostly of Na or K and noble metals (however, still not with
  all particularities)\cite{brack,ekardt}. In this range of cluster dimensions
   the experimentally observed red-shift of Mie frequency  is described mainly
    by spill-out effect, which is pronounced in small nanocrystals and causes a red-shift due to reducing of
     electron density (Mie frequency is proportional to the square of the density)\cite{kresin,ekardt,weick}. But
     for dimensions above only few nanometers, the lowering  of electron density caused by spill-out is of order
      of single percent or smaller \cite{kresin}, and thus for higher nanosphere dimensions (of order of few tens  nm)
       the spill-out  effect is rather unimportant (the resonance frequency shift  caused by spill-out is
        proportional to inverse radius of a metallic sphere, as being typical surface-type effect, and thus diminishes with radius growth).
         The other quantum effects,  as magic electron numbers, related with closed shells for confined system\cite{kresin,ekardt}
          correspond also to small clusters, starting  from only several electrons to few hundreds of electrons (though supershell
         beats, up to $N=3000$ in alkali clusters, were also investigated---for review cf. Ref. \onlinecite{brack}).
           For ultra-small clusters electron collective excitations are strongly coupled between volume and surface and also with background ionic
            system oscillations. Emerging of a well formed volume plasmon for $\sim 0.5$ nm radius was demonstrated\cite{brack,ekardt} and it
             was argued that for smaller clusters separation of surface and volume excitations is impossible (calculations are ranged
             to l=1 dipole mode \cite{ekardt}). With growth of the cluster radius the situation changes and both types of collective excitations:
              translational  and compressional, addressed to surface and volume oscillations, respectively, can be separated (finally they have strongly
              different frequencies). Note, that useful notions of volume and surface oscillations were also applied in the case of nuclear matter vibrations
 of compressional type\cite{migdal,steinw}  and of translational type\cite{goldb}, respectively, which was then
 important in understanding of  giant nuclear dipole resonance experiments.

               The Landau damping due to decay of plasmons for particle-hole pairs with similar as plasmon energy
               was also analyzed (in order to explain observed experimentally resonance frequency red-shift for small clusters)
                which is, however, also a diminishing effect with growing radius \cite{ekardt1,gerch}. Decay of surface plasmons into
                 particle-hole pairs (Landau damping)  leads to red-shift of Mie frequency  (not monotonic with radius for ultra-small clusters)
                  \cite{ekardt1} and similar in scaling, $\sim \frac{1}{a}$, to spill-out-caused shift,   for larger spheres \cite{ekardt1,weick1}
                   (Landau damping is rather limited to radius range up to 2.5 nm)\cite{weick}.

 In the present paper we develop the RPA semiclassical method, originally formulated by Bohm and Pines for bulk metal\cite{rpa,pines},
   in order to describe electron collective excitations in  large metallic nanosphere,
   with radius range 10--60 nm (much larger than Thomas-Fermi radius being of order of interparticle separation),
  including both volume and  surface types of plasmons in the framework of an
all-analytical calculus. The metal is assumed as a so-called 'simple metal', i.e. allowing
for description of the electron-ion interaction by local and not strong pseudopotential (this condition  is satisfied e.g.,
for noble, alkali, or transition metals)\cite{rpa}.
   In the next section, the RPA equations for a local
electron density are derived, including conditions imposed by  finite geometry of the nanosphere (particularities of the solution method are
shifted to  Appendix \ref{app1}). In the following section  spectra
of surface and volume plasmons for the metallic nanosphere are presented, along with  their modification by dielectric medium in which a metallic sphere can be embedded
(the surface plasmon frequencies exhibit  significant dependence on the dielectric constant of  a surrounding material in contrary
to the volume plasmon frequencies). Within this semiclassical RPA attitude, the influence of the volume excitations onto the surface ones is also
described, including possible exciting of surface plasmons by volume electron fluctuations.
Finally, the e-m response of dielectric medium with metallic nanosphere subsystem is analyzed in the case
of the dipole type ($l=1$) excitation, including modifications caused by plasmon damping, resulting in a strong dependence of
 resonance energy shift on a nanosphere radius, which was observed experimentally for large nanospheres\cite{wzr2}.
 Various channels of plasmon damping are
considered including radiation losses in far-field and near-field regimes (in Appendices B and C, respectively)
 and the resulting resonant spectrum modification is compared with experimentally measured
dependence of emission and absorption rates with respect to the  sphere radius\cite{wzr2}
and  dielectric coating\cite{wzr4,wzr5}. The giant strengthening of a coupling in a near-field  regime of surface plasmons with  semiconductor substrate is described
in agreement  with the experimental data for enhancement of a photo-current in metallically nanomodified diode system\cite{wzmocn1,wzmocn2,wzr1,wzr2,wzr3,wzr4}.

\section{RPA semiclassical approach  to electron distribution in metallic nanosphere}
\subsection{Derivation of RPA equation for local electron density in a confined geometry}

Let us consider a metallic sphere with a radius $a$
located in vacuum, $\varepsilon = 1,\;\mu =1$. We assume that
the interaction between electrons and ions is described by a local and weak pseudopotential
(this condition corresponds to the so-called 'simple metal' case)\cite{rpa}, as e.g. for noble metals. The Hamiltonian for this system has the form:
\begin{equation}
\hat{H}=-\sum\limits_{\nu =1}^{N} \frac{\hbar^2 \nabla_{\nu}^2}{2M} +\frac{1}{2}\sum\limits_{\nu\neq \nu'}u({\bm R}_{\nu}   -  {\bm R}_{\nu'})
-\sum\limits_{j =1}^{N_e} \frac{\hbar^2 \nabla_{j}^2}{2m}  + \frac{1}{2}\sum\limits_{j\neq j'}\frac{e^2}{|{\bm r}_{j}   -  {\bm r}_{j'}|}
+\sum\limits_{\nu, j} w({\bm R}_{\nu} - {\bm r}_{j}),
\end{equation}
 where ${\bm R}_{\nu}$,  ${\bm r}_{j}$ and $M$, $m$ are positions and masses of ions and electrons, respectively; $N$---number of ions in the sphere,
$N_e=ZN$---number of collective electrons, $u({\bm R}_{\nu}   -  {\bm R}_{\nu'})$ is the interaction of ions (ion is treated as a nucleus with electron core of
closed shells), $ w({\bm R}_{\nu} - {\bm r}_{j})$ is the local pseudopotential  of electron-ion
interaction.
 Assuming the {\it jellium} model\cite{brack,ekart,ksi} one can write for the background ion charge uniformly distributed
over the sphere: $n_e({\bm r})=n_e \Theta(a-r)$  where  $n_e=N_e/V$ and $n_e|e|$---averaged positive charge density, $V=\frac{4 \pi a^3}{3}$---sphere volume,
$\Theta$ is the Heaviside step-function.
Then, neglecting ion dynamics and small electron-ion pseudopotential (shifted by {\it jellium}-electron  interaction),
collective electrons can be described by the Hamiltonian:
\begin{equation}
 \hat{H}_{e} = \sum\limits_{j =1}^{N_e} \left[ -\frac{\hbar^2 \nabla_{j}^2}{2m} -e^2 \int\frac{n_e ({\bm r})d^3 {\bm r}}{|{\bm r}_{j}   -  {\bm r}|}\right]
 +\frac{1}{2}\sum\limits_{j\neq j'}\frac{e^2}{|{\bm r}_{j}-{\bm r}_{j'}|},
\end{equation}
with corresponding electron wave function $ \Psi_e(t)$.

A local electron density can be written as follows\cite{rpa,pines}:
\begin{equation}
\rho({\bm r}, t)=<\Psi_e(t)|\sum\limits_j \delta({\bm r}-{\bm r}_j) |\Psi_e(t)>,
\end{equation}
with the Fourier picture:
\begin{equation}
\tilde{\rho}({\bm k}, t)=\int \rho({\bm r},t) e^{-i{\bm k}\cdot {\bm r}} d^3 r = <\Psi_e(t)|\hat{\rho}({\bm k})|\Psi_e(t)>,
\end{equation}
where the 'operator' $ \hat{\rho} ({\bm k})=\sum\limits_{j}  e^{-i{\bm k}\cdot {\bm r_j}} $.

Using the above notation one can rewrite $\hat{H}_{e}$ in the following form, in an analogy to the bulk case\cite{rpa,pines}:
\begin{equation}
\hat{H}_{e} = \sum\limits_{j =1}^{N_e} \left[ -\frac{\hbar^2 \nabla_{j}^2}{2m}\right] -
\frac{e^2}{(2 \pi)^3} \int d^3 k \tilde{n}_e({\bm k}) \frac{2 \pi}{k^2} \left(\hat{\rho^{+}}({\bm k}) +  \hat{\rho}({\bm k})\right)
 + \frac{e^2}{(2 \pi)^3}\int d^3 k \frac{2 \pi}{k^2}\left[ \hat{\rho^{+}}({\bm k})  \hat{\rho}({\bm k}) -N_e \right],
\end{equation}
where:
$ \tilde{n}_e({\bm k})=\int d^3 r n_e ({\bm r}) e^{-i{\bm k}\cdot {\bm r}}$, $  \frac{4 \pi}{k^2}= \int d^3 r \frac{1}{r} e^{-i{\bm k}\cdot {\bm r}}$.

Utilizing this form of the electron Hamiltonian one can write out:
\begin{equation}
\frac{d^2 \hat{\rho} ({\bm k}) }{dt^2}=\frac{1}{(i\hbar)^2} \left[\left[ \hat{\rho} ({\bm k}),\hat{H}_{e} \right],\hat{H}_{e} \right],
\end{equation}
in the following form:
\begin{equation}
\label{456}
\begin{array}{l}
\frac{d^2 \hat{\rho} ({\bm k}) }{dt^2}= -\sum\limits_{j}e^{-i{\bm k}\cdot {\bm r}_j}\left\{ -\frac{\hbar^2}{m^2}\left( {\bm k}\cdot \nabla_j \right)^2
+ \frac{\hbar^2 k^2}{m^2}i  {\bm k}\cdot \nabla_j +\frac{\hbar^2 k^4}{4 m^2}\right\}\\
-\frac{4\pi e^2}{m (2\pi )^3}\int d^3q  \tilde{n}_e ({\bm q})\frac{{\bm k}\cdot {\bm q}}{q^2}  \hat{\rho}({\bm k}- {\bm q})
-\frac{4\pi e^2}{m (2\pi )^3}\int d^3q  \hat{\rho}({\bm k}- {\bm q})\frac{{\bm k}\cdot {\bm q}}{q^2}  \hat{\rho}( {\bm q}).\\
\end{array}
\end{equation}
If one  takes into account that $\hat{\rho}({\bm k}- {\bm q}) \hat{\rho}( {\bm q})=
\delta\hat{\rho}({\bm k}- {\bm q}) \delta\hat{\rho}( {\bm q}) + \tilde{n}_e({\bm k}- {\bm q}) \delta\hat{\rho}( {\bm q})
+ \delta\hat{\rho}({\bm k}- {\bm q}) \tilde{n}_e( {\bm q}) +  \tilde{n}_e({\bm k}- {\bm q}) \tilde{n}_e( {\bm q})$ and
$  \tilde{n}_e ({\bm q})\hat{\rho}({\bm k}- {\bm q}) =  \tilde{n}_e ({\bm q})\delta\hat{\rho}({\bm k}- {\bm q})
+   \tilde{n}_e ({\bm q})\tilde{n}_e({\bm k}- {\bm q})$, where    $\delta \hat{\rho}({\bm k}) =  \hat{\rho}({\bm k)} -  \tilde{n}_e ({\bm k})$
describes the 'operator' of
 local electron density fluctuations above the uniform distribution, one can rewrite Eq. (\ref{456})
 in the form:
\begin{equation}
\begin{array}{l}
\frac{d^2 \delta \hat{\rho} ({\bm k}) }{dt^2}=-\sum\limits_{j}e^{-i{\bm k}\cdot {\bm r}_j}\left\{ -\frac{\hbar^2}{m^2}\left( {\bm k}\cdot \nabla_j \right)^2
+ \frac{\hbar^2 k^2}{m^2}i  {\bm k}\cdot \nabla_j +\frac{\hbar^2 k^4}{4 m^2}\right\}\\
-\frac{4\pi e^2}{m (2\pi )^3}\int d^3q  \tilde{n}_e ({\bm k}-{\bm q})\frac{{\bm k}\cdot {\bm q}}{q^2} \delta \hat{\rho}( {\bm q})
-\frac{4\pi e^2}{m (2\pi )^3}\int d^3q \delta \hat{\rho}({\bm k}- {\bm q})\frac{{\bm k}\cdot {\bm q}}{q^2}  \delta \hat{\rho}( {\bm q}).\\
\end{array}
\end{equation}
Thus for the electron density fluctuation,
$\delta \tilde{\rho}({\bm k},t)=   <\Psi_e|\delta\hat{\rho}({\bm k},t)|\Psi_e>=  \tilde{\rho}({\bm k},t) -  \tilde{n}_e ({\bm k})$,
we find:
\begin{equation}
\label{e10}
\begin{array}{l}
\frac{\partial^2 \delta \tilde{\rho} ({\bm k},t) }{\partial t^2}=-<\Psi_e|\sum\limits_{j}e^{-i{\bm k}\cdot {\bm r}_j}\left\{ -\frac{\hbar^2}{m^2}\left( {\bm k}\cdot \nabla_j \right)^2
+ \frac{\hbar^2 k^2}{m^2}i  {\bm k}\cdot \nabla_j +\frac{\hbar^2 k^4}{4 m^2}\right\}|\Psi_e>\\
-\frac{4\pi e^2}{m (2\pi )^3}\int d^3q  \tilde{n}_e ({\bm k}-{\bm q})\frac{{\bm k}\cdot {\bm q}}{q^2} \delta \tilde{\rho}( {\bm q},t)
-\frac{4\pi e^2}{m (2\pi )^3}\int d^3q \frac{{\bm k}\cdot {\bm q}}{q^2} <\Psi_e| \delta \hat{\rho}({\bm k}- {\bm q})  \delta \hat{\rho}( {\bm q})    |\Psi_e>.\\
\end{array}
\end{equation}
Within semiclassical approximation three components of the first term in right-hand-side of Eq. (\ref{e10}) can be estimated as:
 $k^2 v_F^2  \delta \tilde{\rho}({\bm k},t)$, $k^3 v_F/k_T\delta \tilde{\rho}({\bm k},t)$  and
 $ k^4 v_F^2/k_T^2\delta \tilde{\rho}({\bm k},t)$, respectively ($1/k_T$ is  Thomas-Fermi  radius\cite{rpa}, $k_T=\sqrt{\frac{6\pi n_e e^2}{\epsilon_F}}$,
 $\epsilon_F$---Fermi energy, $v_F$---Fermi velocity).
 The contributions of the second and the third components of the first term can be neglected in comparison to the first component. Small and thus negligible  is also
the  third term  in right-hand-side of Eq.(\ref{e10}), as involving a product of two $ \delta \tilde{\rho}$ (which we assumed small, $ \delta \tilde{\rho}/n_e << 1$).
This approach corresponds to random-phase-approximation (RPA) attitude formulated for bulk metal\cite{rpa,pines}
(note that $\delta \hat{\rho}(0)=0$ and the coherent RPA contribution
of interaction is comprised by the last but one term in Eq. (\ref{e10})).

Within the RPA   Eq. (\ref{e10}) attains thus the form:
\begin{equation}
\label{e11}
\frac{\partial^2 \delta \tilde{\rho} ({\bm k},t) }{\partial t^2}=\frac{2 k^2}{3m}
<\Psi_e|\sum\limits_{j}e^{-i{\bm k}\cdot {\bm r}_j}\frac{\hbar^2\nabla_j^2}{2m}|\Psi_e>
-\frac{4\pi e^2}{m (2\pi )^3}\int d^3q  \tilde{n}_e ({\bm k}-{\bm q})\frac{{\bm k}\cdot {\bm q}}{q^2} \delta \tilde{\rho}( {\bm q},t),
\end{equation}
where for the case of spherical symmetry:
$$ <\Psi_e|\sum\limits_{j}e^{-i{\bm k}\cdot {\bm r}_j}\frac{\hbar^2}{m^2}\left( {\bm k}\cdot \nabla_j \right)^2|\Psi_e>
\simeq \frac{2 k^2}{3m}<\Psi_e|\sum\limits_{j}e^{-i{\bm k}\cdot {\bm r}_j}\frac{\hbar^2\nabla_j^2}{2m}|\Psi_e>. $$
In the position representation Eq. (\ref{e11}) can be rewritten in  the following manner:
\begin{equation}
\label{e12}
\begin{array}{l}
\frac{\partial^2 \delta \tilde{\rho} ({\bm r},t) }{\partial t^2}=-\frac{2 }{3m} \nabla^2
<\Psi_e|\sum\limits_{j}\delta({\bm r}-{\bm r}_j)\frac{\hbar^2\nabla_j^2}{2m}|\Psi_e>\\
+\frac{\omega_p^2}{4\pi} \nabla \left\{ \Theta(a-r) \nabla \int d^3r_1 \frac{1}{|{\bm r}-{\bm r}_1|} \delta \tilde{\rho}( {\bm r}_1,t)\right\}.\\
\end{array}
\end{equation}
The Thomas-Fermi  averaged kinetic energy can be represented as follows\cite{rpa}:
\begin{equation}
\begin{array}{l}
<\Psi_e|-\sum\limits_{j}\delta({\bm r}-{\bm r}_j)\frac{\hbar^2\nabla_j^2}{2m}|\Psi_e>\simeq
\frac{3}{5} (3\pi^2)^{2/3} \frac{\hbar^2}{2m} (\rho({\bm r},t))^{5/3}\\
=\frac{3}{5} (3\pi^2)^{2/3} \frac{\hbar^2}{2m}n_e^{5/3} \Theta(a-r)\left[1+\frac{5}{3}\frac{\delta \tilde{\rho}({\bm r},t)}{n_e}+...\right].\\
\end{array}
\end{equation}
Note, that neglected here gradient terms (in particular von Weizs\"acker term, $\sim (\nabla \rho)^2/(4\rho)$, beyond the
 Thomas-Fermi formula for kinetic energy functional  $\sim  \rho^{5/3} $)\cite{brack}
 strongly affect the finite system properties especially of small metallic clusters. The contribution of this particular
 term (von Weizs\"acker) depends on the approximation in various versions of corrections to Thomas-Fermi approach\cite{kresin}
 (the coefficient of   von Weizs\"acker term  is treated  even as a convenient fitting parameter).
 The gradient terms are inexplicitly  included in TDLDA type methods  based on Kohn-Sham equation.
 As it follows from respective analyses the contributions related to these terms (mostly spill-out effect)
  are more important for small clusters (when the surface dominates)  and gradually diminish with the nanosphere
   radius growth \cite{brack,kresin,ekardt,weick,weick1}.

Taking then into account that $\nabla \Theta(a-r)=-\frac{\bm r}{r}\delta(a-r)= -\frac{\bm r}{r}\lim_{\epsilon\rightarrow 0}\delta(a+\epsilon-r)$,
one can rewrite Eq. (\ref{e12}) in the following manner:
\begin{equation}
\label{e15}
\begin{array}{l}
\frac{\partial^2 \delta \tilde{\rho} ({\bm r},t) }{\partial t^2}=\left[ \frac{2}{3} \frac{\epsilon_F}{m}\nabla^2 \delta \tilde{\rho}( {\bm r},t)-
\omega_p^2 \delta \tilde{\rho}( {\bm r},t)\right]\Theta(a-r)\\
- \frac{2}{3m} \nabla\left\{\left[\frac{3}{5}\epsilon_F n_e+\epsilon_F \delta \tilde{\rho}( {\bm r},t)\right]\frac{\bm r}{r}\delta(a+\epsilon-r)
\right\}\\
-\left[\frac{2}{3} \frac{\epsilon_F}{m}\frac{\bm r}{r}\nabla \delta \tilde{\rho}( {\bm r},t)
+      \frac{\omega_p^2}{4\pi}          \frac{\bm r}{r}\nabla \int d^3r_1 \frac{1}{|{\bm r}-{\bm r}_1|} \delta \tilde{\rho}( {\bm r}_1 ,t)\right]
\delta(a+\epsilon-r).\\
\end{array}
\end{equation}
In the above formula  $\omega_p$ is the bulk plasmon frequency,  $\omega_p^2=\frac{4\pi n_e e^2}{m}$, and
      the abbreviated notation,  $\delta(a+\epsilon-r)   =  \lim_{\epsilon\rightarrow 0}\delta(a+\epsilon-r)$, was used.
The solution of Eq. (\ref{e15}) can be decomposed into two parts  related   to  the distinct domains:
\begin{equation}
 \delta \tilde{\rho}( {\bm r,t})=\left\{
           \begin{array}{l}
             \delta \tilde{\rho}_1( {\bm r,t}), \;for\; r<a,\\
              \delta \tilde{\rho}_2( {\bm r,t}), \;for\; r\geq a,\; ( r\rightarrow a+),\\
          \end{array}
       \right.
       \end{equation}
corresponding to the volume and surface  excitations, respectively. These two parts of local electron density fluctuations
satisfy the equations:
\begin{equation}
\label{e20}
\frac{\partial^2 \delta \tilde{\rho}_1 ({\bm r},t) }{\partial t^2}=\frac{2}{3} \frac{\epsilon_F}{m}\nabla^2 \delta \tilde{\rho}_1( {\bm r},t)-
\omega_p^2 \delta \tilde{\rho}_1( {\bm r},t),
\end{equation}
and
\begin{equation}
\label{e21}
\begin{array}{l}
\frac{\partial^2 \delta \tilde{\rho}_2 ({\bm r},t) }{\partial t^2} =-
\frac{2}{3m} \nabla\left\{\left[\frac{3}{5}\epsilon_F n_e+\epsilon_F \delta \tilde{\rho}_2( {\bm r},t)\right]\frac{\bm r}{r}\delta(a+\epsilon-r)\right\}\\
 -  \left[\frac{2}{3} \frac{\epsilon_F}{m}\frac{\bm r}{r}\nabla \delta \tilde{\rho}_2( {\bm r},t)
+      \frac{\omega_p^2}{4\pi}          \frac{\bm r}{r}\nabla \int d^3r_1 \frac{1}{|{\bm r}-{\bm r}_1|} \left(\delta \tilde{\rho}_1( {\bm r}_1 ,t)
\Theta(a-r_1)+\delta \tilde{\rho}_2( {\bm r}_1 ,t)\Theta(r_1-a)\right)\right]\delta(a+\epsilon-r).\\
\end{array}
\end{equation}

Within this quasiclassical simplified approach volume plasmons  described  by the Eq. (\ref{e20}) are independent of the surface plasmons, while the latter
can be excited by the former ones, due to the last term in Eq. (\ref{e21}) (which is caused by a 'surface tail' of
compressional---volume type oscillations, while oppositely, translational---surface type
 oscillations do not have a 'volume tail'), which expresses a coupling between surface and volume plasmons
in large metallic nanosphere.
Coupling between volume and surface plasmons was
 analyzed in TDLDA approach for jellium spheres \cite{brack,ekardt}
  and  it was demonstrated that for ultra-small clusters it is impossible to decouple
   volume and surface oscillations (because of perturbation of a shell structure), while for bigger
    clusters (more than 58 electrons, for Na cluster)  the well formed and separated both modes emerge\cite{brack,ekardt}
     (the TDLDA  analyses were done for clusters up to ca 200 electrons). This volume--surface coupling is strong for
     ultra-small radii, when shell effects and quantum spill-out   are pronounced, and gradually weakens with growing sphere dimension.

     In the present paper we consider the
       radius range 10--60 nm, when  quantum effects, significant for smaller clusters, are not of primary importance
        and are dominated by irradiation behavior.
Such large nanospheres contain $10^5$--$10^7$ collective electrons,
  thus are considerably larger than clusters with up to 200 electrons, numerically
  investigated in details by use of  TDLDA methods. Our main idea is to formulate analytical
  (thus simplified) RPA description in the form of  oscillator equation allowing for phenomenological
   inclusion of damping rates due to irradiation losses dominating surface oscillation behavior
    in the considered range of nanosphere dimension. Next we apply such a model
    to explanation of experimentally observed giant increase in photo-effect efficiency
    due to near-field coupling of  surface plasmons with electrons in substrate semiconductor,
    when large metallic nanospheres are deposited on the optically active  photo-diode layer.
     This coupling creates very effective channel for energy transfer as it will be demonstrated below.

\subsection{Solution of RPA equations: volume and surface plasmon frequencies}

Eqs (\ref{e20}) and (\ref{e21}) can be solved upon imposed  boundary and symmetry conditions---cf. Appendix \ref{app1}.
Let us represent  both parts of the electron fluctuation in the following manner:
\begin{equation}
           \begin{array}{l}
             \delta \tilde{\rho}_1( {\bm r,t})=n_e\left[f_1(r)+F({\bm r}, t)\right], \;for\; r<a,\\
              \delta \tilde{\rho}_2( {\bm r,t})=n_e f_2(r)+\sigma(\Omega,t)\delta(r+\epsilon -a), \;for\; r\geq a,\; ( r\rightarrow a+),\\
          \end{array}
       \end{equation}
and now let us choose the convenient initial conditions, $  F({\bm r}, t)|_{t=0}=0, \; \sigma(\Omega,t)|_{t=0}=0$, ($\Omega$---spherical angle), moreover
$(1+f_1(r))|_{r=a}=f_2(r)|_{r=a}$ (continuity condition), $ F({\bm r}, t)|_{r=a}=0$, $\int\rho({\bm r},t)d^3r=N_e$ (neutrality condition).

We arrive thus with the explicit form of the solutions of   Eqs (\ref{e20}, \ref{e21}) (as it is described in  Appendix \ref{app1}):
\begin{equation}
\begin{array}{l}
f_1(r)=-\frac{k_T a +1}{2} e^{-k_T (a-r)} \frac{1-e^{-2k_Tr}}{k_Tr}, \; for \;\;r<a,\\
f_2(r)=\left[k_Ta - \frac{k_Ta+1}{2}\left(1-e^{-2k_Ta}\right)\right]\frac{e^{-k_T(r-a)}}{k_Tr}, \; for\;\; r\geq a,\\
\end{array}
\end{equation}
where $k_T=\sqrt{\frac{6\pi n_e e^2}{\epsilon_F}}=\sqrt{\frac{3\omega_p^2}{v_F^2}}$.
For time-dependent parts of electron fluctuations we find:
\begin{equation}
\label{e2001}
F({\bm r}, t) =\sum\limits_{l=1}^{\infty}\sum\limits_{m=-l}^{l}\sum\limits_{n=1}^{\infty}A_{lmn}j_{l}(k_{nl}r)Y_{lm}(\Omega)sin(\omega_{nl}t),
\end{equation}
and
\begin{equation}
\label{e25}
\begin{array}{l}
\sigma(\Omega,t)   = \sum\limits_{l=1}^{\infty}\sum\limits_{m=-l}^{l}  \frac{B_{lm}}{a^2}Y_{lm}(\Omega)sin(\omega_{0l}t)\\
+  \sum\limits_{l=1}^{\infty}\sum\limits_{m=-l}^{l}\sum\limits_{n=1}^{\infty}
A_{lmn}\frac{(l+1)\omega_p^2}{l\omega_p^2-(2l+1)\omega_{nl}^2}Y_{lm}(\Omega)n_e\int\limits_0^a dr_1 \frac{r_1^{l+2}}{a^{l+2}}j_{l}(k_{nl}r_1)sin(\omega_{nl}t),\\
\end{array}
\end{equation}
where $j_l(\xi)=\sqrt{\frac{\pi}{2\xi}}I_{l+1/2}(\xi)$---the spherical Bessel function, $Y_{lm}(\Omega)$---the spherical function, $\omega_{nl}=
\omega_p\sqrt{1+\frac{x_{nl}^2}{k_T^2a^2}}$---the frequencies of electron volume self-oscillations (volume plasmon frequencies),
$x_{nl}$---nodes of the Bessel function $j_l(\xi)$,  $k_{nl}=x_{nl}/a$,
 $\omega_{0l}=\omega_p\sqrt{\frac{l}{2l+1}}$---the frequencies of electron surface self-oscillations (surface plasmon frequencies).

From the above  it follows thus that  local electron density (within RPA attitude) has the form:
\begin{equation}
\label{e50}
\rho({\bm r},t)=\rho_0(r)+\rho_{neq}({\bm r},t),
\end{equation}
where the  RPA  equilibrium electron distribution (correcting the uniform distribution $n_e$):
\begin{equation}
  \rho_0(r)=\left\{
  \begin{array}{l}
  n_e\left[1+f_1(r)\right],\; for\;\; r<a,\\
  n_ef_2(r),\;for\;\; r\geq a, \; r\rightarrow a+\\
  \end{array} \right.
  \end{equation}
and the nonequilibrium, of plasmon oscillation type:
\begin{equation}
  \rho_{neq}({\bm r},t)=\left\{
  \begin{array}{l}
  n_eF({\bm r},t),\; for\;\; r<a,\\
  \sigma(\Omega,t)\delta(a+\epsilon-r)\;for\;\; r\geq a,\; r\rightarrow a+.\\
  \end{array} \right.
  \end{equation}
 The function $F({\bm r},t)$ displays volume plasmon oscillations, while
  $ \sigma(\Omega,t)$ describes the surface plasmon oscillations. Let us emphasize that
  in the formula for   $ \sigma(\Omega,t)$, Eq. (\ref{e25}), the first term corresponds to surface self-oscillations,
  while the second term describes the surface  oscillations induced by the volume plasmons.
  The frequencies of the surface self-oscillations are
  \begin{equation}
  \omega_{0l}=\omega_p\sqrt{\frac{l}{2l+1}},
  \end{equation}
  which, for $l=1$, agrees with the dipole type surface oscillations described originally by Mie\cite{mie}, $\omega_{01}=\omega_p/\sqrt{3}$,

\subsection{Surface plasmon frequencies for metallic nanosphere embedded in a dielectric medium, with $\varepsilon > 1$}

In order to account for influence of dielectric surroundings  on the surface plasmons in
metallic nanosphere, let us assume that  electrons on the surface ($r=a+$, i.e. $r\geq a,\;r\rightarrow a$)
interact with Coulomb forces renormalized  by the relative dielectric constant
$\varepsilon > 1$. Thus instead of Eq. ({\ref{e21})  one can  consider its following
modification:
\begin{equation}
\label{e26}
\begin{array}{l}
\frac{\partial^2 \delta \tilde{\rho}_2 ({\bm r}) }{\partial t^2} =-
\frac{2}{3m} \nabla\left\{\left[\frac{3}{5}\epsilon_F n_e+\epsilon_F \delta \tilde{\rho}_2( {\bm r},t)\right]\frac{\bm r}{r}\delta(a+\epsilon-r)\right\}\\
 -  \left[\frac{2}{3} \frac{\epsilon_F}{m}\frac{\bm r}{r}\nabla \delta \tilde{\rho}_2( {\bm r},t)
+      \frac{\omega_p^2}{4\pi}          \frac{\bm r}{r}\nabla \int d^3r_1 \frac{1}{|{\bm r}-{\bm r}_1|} \left(\delta \tilde{\rho}_1( {\bm r}_1 ,t)
\Theta(a-r_1)+\frac{1}{\varepsilon}\delta \tilde{\rho}_2( {\bm r}_1 ,t)\Theta(r_1-a)\right)\right]\delta(a+\epsilon-r),\\
\end{array}
\end{equation}
(Eq. (\ref{e20}) remains in not changed form).
Solution of the above equation is of the similar form as that for the Eq. (\ref{e21}) case, but with
new surface plasmon frequencies:
\begin{equation}
 \omega_{0l}=\omega_p\sqrt{\frac{l}{2l+1}\frac{1}{\varepsilon}}.
  \end{equation}
The frequency of surface electron self-oscillations, changed by the factor $\sqrt{\frac{1}{\varepsilon}}$,
can be reduced significantly  in comparison
 to the vacuum case, as
in many materials $\varepsilon$ is relatively big ($\varepsilon$ corresponds to its  high-frequency limiting value, the same which is
involved in  a refraction coefficient).

Our result for resonant surface plasmon frequency,
 $\omega_p\sqrt{\frac{l}{\varepsilon (2l+1)}}$,
does not reproduce, for dipole case $l=1$, the classical Mie formula\cite{rubio,petrov},
 $\omega_p\frac{1}{\sqrt{2\varepsilon+1}}$.
 Our frequency is lower than the Mie one, which corresponds well
 with the data indicated in  Fig. 3  in Ref. \onlinecite{rubio},
 in which there are presented resonance frequencies obtained within a more thorough (TDLDA)
  method and located also below corresponding classical  Mie values for several
   dielectric constants of surrounding medium ($\varepsilon =1\; 1.7, \; 1.95,\; 2.31,\; 3$ for air, Ar, Kr, Xe, MgO, respectively).
   Our formula has the similar property as includes some quantum effects (RPA approach) in comparison to completely classical Mie derivation\cite{petrov}.

\section{Evaluation of a damping rate for surface plasmons}

The RPA semiclassical  equations (\ref{e20}, \ref{e21}) for plasmon excitations   reveal the form of oscillator-equation-type. Thus it is easy to include, in the phenomenological
manner, attenuation of these excitations, via  damping term $\frac{2}{\tau^{(i)}}\frac{\partial \rho_{i}({\bm r},t)}{\partial t}$ which can be
added to the left-hand-side of Eq. (\ref{e20}) and Eq. (\ref{e21}) (assuming that  the volume modes, $i=1$, and
the surface modes, $i=2$, are damped with the  attenuation times $\tau^{(i)}$, respectively).
Thus the time-dependent solution of such modified equation (\ref{e20}) attains the form as given by Eq. (\ref{e2001})  with the factor
$e^{-t/\tau^{(1)}}$, and shifted frequency $\omega'_{nl}=\sqrt{\omega_{nl}^2-\frac{1}{(\tau^{(1)})^2}}$ for the volume modes. Similarly  for
the surface plasmons  [Eq. (\ref{e21})], the attenuation leads to the factor $e^{-t/\tau^{(2)}}$ for the first part of the solution
(\ref{e25}) (and simultaneously shifted frequency   $\omega'_{0l}=\sqrt{\omega_{0l}^2-\frac{1}{(\tau^{(2)})^2}}$), while the second term of
Eq. (\ref{e25}) acquires an additional factor $e^{-t/\tau^{(1)}}$  (and shifted frequency  $\omega'_{nl}=\sqrt{\omega_{nl}^2-\frac{1}{(\tau^{(1)})^2}}$).

There are various mechanisms of energy dissipation of plasmon oscillations in metallic nanospheres. Let us concentrate on damping of surface plasmons
described by $\frac{1}{\tau^{(2)}}$.  The interaction with phonons,  electrons and
lattice defects contribute to the relaxation rate $\frac{1}{\tau^{(2)}}$ with the  $\frac{1}{\tau^{(2)}_1}$, which is determined
by the mean free path of electrons in the nanosphere, reduced additionally in comparison to the bulk case by inelastic scatterings with the sphere surface.
One can use the estimation\cite{atwater} $\frac{1}{\tau^{(2)}_1}\sim \frac{v_F}{\lambda_B}+\frac{C v_F}{a}$, where $v_F$---Fermi velocity,
$\lambda_B$---an effective value of the mean free path, $C$---constant of order 1, $a$---nanosphere radius (for Ag, $v_F=1.4\cdot 10^6$ m/s, $\lambda_B\simeq 57$ nm,
which for $a=25$ nm gives  $\frac{1}{\tau^{(2)}_1}=8\cdot10^{13}$ s$^{-1}$, while rather a femtosecond decay time agrees
with the measurements on Ag nanoparticles\cite{ag}).
Note that decomposition of surface plasmons due to creation of particle--hole pairs (Landau damping)\cite{ekardt1,weick1} is efficient only
for small clusters\cite{weick1}.

Another type of energy dissipation can be associated with the radiation decay.
The far-field radiation (i.e. for distances much longer than the wave-length $\lambda \gg a$)
gives the contribution to the relaxation $\frac{2}{\tau^{(2)}_2}\sim \frac{2e^2}{3mc^3}\omega_1^2\sim 1.6\cdot10^{8}$ s$^{-1}$
(for $\omega_1\sim 5\cdot 10^{15}$ s$^{-1}$)
per single electron due to the Lorentz
friction\cite{lan}. If one multiplies it by the electron number  $N_e=\frac{4\pi a^3}{3}n_e, \;\;  n_e=\frac{m\omega_p^2}{4\pi e^2}$  (in order to account for
the probability of energy transfer from the total system),
one can arrive at the value  $\frac{1}{\tau^{(2)}_2} = \omega_{1} \frac{1}{3} \left(\frac{\omega_p a}{\sqrt{3}c}\right)^3$, which dominates  over $\frac{1}{\tau^{(2)}_1}$
for not too small spheres\cite{ksi,kre,ag}.
This
channel of plasmon energy dissipation allows for explanation  of the surface plasmon oscillations  behavior with growing $a$ (as  $\frac{1}{\tau^{(2)}_2}$
scales as $a^3$)
for  nanospheres embedded in a dielectric medium (like in the water---as it is  presented in the Tab. 1 for nanoparticles of gold).
The more precise derivation of the far-field radiation losses expressed by   $\frac{1}{\tau^{(2)}_2}$ is presented in  Appendix \ref{app2}, leading  to
the same formula for  $\frac{1}{\tau^{(2)}_2}$  as  given above.
Note that if the attenuation rate  $\omega_1\tau^{(2)}_2$ is closer to 1 then the  attenuation induced shift of
the self-frequency is greater, $\omega'=\omega_1\sqrt{1-(\omega_1\tau^{(2)}_2)^{-2}}$ and   this behavior  coincides with experimental observations\cite{wzr1,wzr2}
 (for $\frac{1}{\omega_1 \tau^{(2)}_2}\geq 1$ the overdamped regime is
attained without free plasmon oscillations).

\begin{tabular}{|p{6cm}|p{4cm}|p{1.5cm}|p{1.5cm}|p{1.5cm}|}
\multicolumn{5}{c}{Tab. 1.   Comparison with experimental data\cite{wzr2} for Au nanospheres in  water \label{table1}}\\
\hline
nanosphere radius &$a$& 50 nm& 40 nm& 25 nm \\
\hline
attenuation rate due to far-field radiation losses  & $\omega_1 \tau^{(2)}_2$&1.51&2.95&12.09\\
\hline
shifted self-frequency rate  & $\frac{\omega'}{\omega_1}=\sqrt{1-(\omega_1\tau^{(2)}_2)^{-2}}$&0.75&0.94&0.99\\
\hline
red-shifted oscillation energy &$\hbar\omega'$ (theor.)&2.16 eV&2.70 eV&2.87 eV\\
\hline
red-shifted oscillation energy &$\hbar\omega'$ (exper.)&2.16 eV&2.26 eV&2.36 eV\\
\hline
\end{tabular}

\vspace{3mm}

For Au, $\hbar\omega_p=8.57$ eV, and surface plasmon energy $\hbar\omega_1=2.87$ eV.
This value of  $\hbar\omega_1$ is estimated assuming a coincidence of theoretically predicted self-frequency shifted by attenuation
with experimentally measured   for $a=50$ nm; note that the discrepancy
between  the experimental red-shift and the
 theoretical one grows for smaller $a$, which is probably caused by the strengthening of an impact of $\frac{1}{\tau^{(2)}_1}\sim 1/a$ at smaller $a$, resulting
 in decrease of the red-shift in comparison to  its value caused by  $\frac{1}{\tau^{(2)}_2}\sim a^3$. This tendency at decreasing radius $a$
 seems to be confirmed also by measurements for silver clusters with small dimensions $\leq 10$ nm, which was reported in Refs \onlinecite{stietz,scharte}.

Note that for small metallic clusters,  quantum
 spill-out of electron cloud beyond positive jellium   causes a red-shift of resonance Mie frequency lowering, however,
  with radius growth (thus it is  in fact a blue-shift with radius growth).
    The additional effect of polarization of ionic system (this effect is beyond jellium model)
     can lead oppositely  to inverse  frequency shift, though rather small.
      Some jellium oscillation corrections can  be also accounted for  as scattering
       with phonons and can be included  in the effective time rate for damping
        via effective mean free path $\lambda_B$. The corresponding  contributions are, however,
        significant  rather for small systems \cite{brack,ekardt} (for summarizing of various effects caused red and blue shifts of
        resonance with radius growth, cf. also Ref. \onlinecite{kreib}). We consider the range of sphere dimensions
         when radiation losses cause overwhelming contribution to damping and to the resulting red-shift of the surface plasmon resonance,
          cf. Fig. 1. In this figure it is presented comparison
          of damping contributions due to scattering effects, $\sim \frac{v_F}{\lambda_B} +\frac{v_F}{a}$
          and due to radiation  losses in dielectric surroundings, $\sim a^3$. For $a>10$ nm the latter channel clearly  dominates.
            The radiation-caused red-shift  grows strongly  with the radius of the nanosphere similarly as it is
           observed in the experiment for  range of sphere radii 25--50 nm (for Au) \cite{wzr2}.

The next source of the attenuation of surface plasmons would be connected with the transport of
dipole oscillation energy between nanoparticles due to the F{\"o}rster-type coupling\cite{atwater} in the case of
sufficiently dense location of metallic nanoparticles.
Nevertheless, taking into account that for uniform nanoparticle distribution in the dielectric medium, the same energy rates simultaneously escape and arrive
at
particular nanosphere due to interaction with other nanospheres (nearest-neighbors),  this coupling  does not contribute to the relaxation time
(at least for uniformly distributed metallic nanocomponents).

The situation changes, however, significantly if  metallic nanoparticles are deposited on the surface of the semiconductor substrate.
Then the near-field  e-m energy
transfer from oscillating dipoles (surface plasmons with $l=1$) to the electrons in substrate semiconductor starts
 to be the dominant channel of surface plasmon dissipation.
One can estimate the corresponding time-rate by the Fermi golden rule applied to the system  of plasmons coupled in near-field zone with
semiconductor substrate. One can consider two situations, the first one---with rapidly switched off  external electric field, which excites
surface plasmons,  then  gradually (with lowering amplitude of oscillations) transferring energy to the semiconductor, and the second one---a stationary state
of plasmons (with  constant amplitude) with mediating role of plasmons transferring  entire energy of incident photons to semiconductor. The latter case
corresponds thus
to a stationary solution of a driven and damped oscillator, while the former one to free damped oscillations. In both cases the
damping rate is the same, as it corresponds to the same substrate in a near-field zone.  For free damped oscillations the total
initial oscillation energy (assessed in  Appendix \ref{app2}) is gradually lost with the time ratio $\frac{1}{\tau^{(2)}_3}$. It allows
for calculation of the  $\frac{1}{\tau^{(2)}_3}$, which is presented in the Appendix \ref{app3}.
Utilizing  the similar calculus as in Appendix \ref{app2}
one can assess the value of the corresponding damping rate $\frac{1}{\tau^{(2)}_3}$, assuming that the   total energy loss of
surface plasmons is transferred to the semiconductor substrate with additional renormalization by a factor $\beta$
lowering an efficiency of this channel ($\beta$ is a phenomenological factor introduced
in order to account for geometry-induced proximity type constraints imposed on the dipole near-field coupling of the nanosphere with
underlying semiconductor layer). Thus it is sufficient to calculate the
energy income in the semiconductor due to nanosphere near-field dipole coupling---as it is done
in  Appendix \ref{app3} (within the Fermi golden rule scheme). For this channel of surface plasmon energy dissipation we deal
with the  scaling   of the resonance energy shift with the dot radius, similar as  that for  $\frac{1}{\tau^{(2)}_2}$, however,
 with possible correction induced by $\beta$  dependence
on $a$ (it may be important, as for the nanosphere located on the planar semiconductor surface one can use
an estimation  $\beta\sim c \frac{h^2}{a^2}\sim 10^{-3}$, [for $a=50$ nm]  where $c$ is a constant, $h$ is an effective range
of near-field coupling).
The parameter $\beta$ significantly  grows  in the case when the total nanosphere is in the near-field contact with the substrate,  i.e. when the nanosphere is
completely embedded in the semiconductor medium.
For nanospheres deposited on the real semiconductor surface, the parameter $\beta$ is obtained through fitting  the experimental data
(cf. Tab. 2).

Assuming stationary conditions (i.e., constant in time amplitude of the surface plasmon oscillations, which corresponds to a balance of
the  incoming   energy of incident photons with the energy outgoing to semiconductor substrate) the  relevant damping  is governed by the
near-field dipole interaction (for $R\ll \lambda$) expressed  by the scalar potential\cite{lan} with an  amplitude $D_0(\omega)$,
\begin{equation}
\varphi({\bm R},t)=\frac{1}{\varepsilon_0 R^2} {\bm n}\cdot {\bm D_0}(\omega)sin(\omega t).
\end{equation}
The matrix element of near-field dipole interaction for the transition of a semiconductor electron from
the state in the valence to the conduction band, assumed as $\Psi_{{{i(f)}}}({\bm r},t)=(2\pi)^{-3/2}exp\left
[i{\bm k}\cdot{\bm r}-iE_{i(f)}({\bm k})t/\hbar\right]$ ($i$--initial, $f$--final, respectively) is
calculated in  Appendix \ref{app3}, (Eq. (\ref{5000})), which leads to a probability of transition per time unit,
$\delta w =\frac{e^2(D_0(\omega))^2 \mu \sqrt{m_p^*m_n^*}}{3(4\pi^3)^2\hbar^5\varepsilon^2} (\hbar\omega -E_g)$,
where $D_0(\omega)$ is the surface plasmon dipole oscillation amplitude, adjusted  to the balance of energy income and outcome (via
shift of the resonance for stationary driven and damped oscillations).

Taking into account that the number of incident photons in the volume $V$ of semiconductor equals
$\frac{\varepsilon E_0^2 V}{8\pi \hbar \omega}$ and the volume rate of metallic components
$C_0=N_m\frac{4\pi a^3}{3V}$ ($N_m$---the number of nanospheres), the probability that an energy of a single incident photon  is transferred to the semiconductor via
surface plasmons on metallic nano-admixtures can be expressed as (with $\delta w$ given by Eq. (\ref{5000})):
\begin{equation}
\label{qm}
q_m=\beta N_m \delta w  \left(\frac{\varepsilon E_0^2 V}{8\pi \hbar \omega}\right)^{-1}=
\frac{\beta C_0e^2\omega f^2(\omega) 4\pi a^3}{128\pi^5\hbar^4\varepsilon} \mu
\sqrt{m_p^*m_n^*}(\hbar\omega -E_g),
\end{equation}
where
$f(\omega) = \frac{\omega_1^2}{\sqrt{(\omega_1^2-\omega^2)^2+4\omega^2/(\tau^{(2)}_3)^2}}$ is the  amplitude of forced
surface plasmon oscillations.

In order to assess an efficiency of the near-field coupling channel
 one can estimate the ratio of probability of energy absorption in semiconductor via mediation of
 surface plasmons (per single photon incident on the
metallic nanospheres)  to the energy attenuation in semiconductor
directly from a planar wave illumination (also per single photon). In the latter case the energy attenuation
in the semiconductor per single incident photon is given by
the formula for ordinary photo-effect,
$q=\frac{2\sqrt{2}}{3\pi^6}\frac{e^2\mu^{5/2}}{m^*_p\omega\varepsilon\hbar^3}\left(\hbar\omega-E_g\right)^{3/2}$ (cf. e.g., Ref. \onlinecite{pol}).
  The  ratio  $\frac{q_m}{q}$ turns out to be
of order of $10^5\frac{\beta 40}{H[nm]}$  (at a typical surface density of nanoparticles, $n_s\sim 10^8$/cm$^2$) which
(including the phenomenological factor $\beta$ and $H$---semiconductor photo-active layer depth) is sufficient to explain the scale of
 experimentally observed strong
enhancement of absorption and emission rates. It should be noticed that  $\frac{1}{\tau_{3}^{(2)}}$ grows  with $\beta$
(cf. Eq. (\ref{tau})) and would attain the critical value for overdamped oscillator ($\frac{1}{\tau_{3}^{(2)}\omega_1}=1$), which
precludes surface plasmon  free oscillations.

Very high efficiency (even if diminished by $\beta$) of the  near-field energy transfer from surface plasmons to semiconductor substrate is caused mainly by a
contribution of all interband transitions, not restricted here to the direct (vertical) ones as for ordinary photo-effect, due to  absence of the momentum
conservation constraints for nanosystems---cf. Appendix \ref{app3}. The strengthening of the probability transition due to all indirect
interband  paths
of excitations  in semiconductor is probably responsible for observed experimentally strong enhancement of
light absorption and emission in diode systems mediated by surface plasmons in nanoparticle
surface coverings\cite{wzmocn1,wzmocn2,wzr1,wzr2,wzr3,wzr4}.

 In the balanced state of the system when the incoming energy of light is transferred to the semiconductor via
 near-field coupling, we deal with the stationary solution of driven and damped oscillator.
          The driving force is the electric field  of the incident planar wave, the
 damping force is the
 near-field energy transfer described by the $\frac{1}{\tau^{(2)}_3}$ (assuming that this dissipation channel is dominating).
  The resulting red-shifted resonance with  simultaneously reduced amplitude
 allows for the accommodation to the balance  of energy transfer to semiconductor with incident photon energy.
 The amplitude of resonant plasmon oscillations  $D_0 (\omega)$ is thus shaped by
 $f(\omega)=\frac{1}{\sqrt{(\omega_1^2-\omega^2)^2+4\omega^2/(\tau^{(2)}_3)^2}}$.
The extremum of red-shifted resonance
is attained at  $\omega_m=\omega_1\sqrt{1-2(\omega_1\tau^{(2)}_3)^{-2}}$ with corresponding amplitude
$\sim \tau^{(2)}_3/\left(2\sqrt{\omega_1^2-(\tau^{(2)}_3)^{-2}}\right)$. This shift is proportional to $1/(\omega_1 (\tau^{(2)}_3)^2)$ and scales
 with nanosphere radius $a$
similarly (diminishes with decreasing $a$) as in the experimental observations\cite{wzr2} (note again that for $1/\tau^{(2)}_1$ the dependence on  $a$ is
opposite [grows with decreasing $a$]).

In order to compare  with the experiment let us estimate the photo-current in the case of metallically modified
 surface in relation to the ordinary photo-effect. The photo-current is given by
 $I'=|e|N(q+q_m)A$, where $N$ is the number of incident photons, $q$ and $q_m$ are probabilities of single photon attenuation
 in ordinary photo-effect\cite{pol} and due to presence of metallic nanospheres, i.e., of
 $q=\frac{2\sqrt{2}}{3\pi^6}\frac{e^2\mu^{5/2}}{m^*_p\omega\varepsilon\hbar^3}\left(\hbar\omega-E_g\right)^{3/2}$ (cf. Ref. \onlinecite{pol})
 and
 $q_m$  given by Eq. (\ref{qm}); $A=\frac{\tau^n_f}{t_n}+\frac{\tau^p_f}{t_p}$ is the
 amplification factor ($\tau^{n(p)}_f$ the annihilation time of both sign carriers, $t_{n(p)}$ the drive time for carriers [the time of traversing the distance between
 electrodes]).
 From the above formulae it follows that  (here $I=I'(q_m=0)$, i.e., the photo-current without metallic modifications),
 \begin{equation}
 \label{fot}
 \frac{I'}{I}
 =1+27.53\cdot  10^6c_0\frac{m^*_p}{m^*_n}\left(\frac{2a}{100[nm]}\sqrt{\frac{\hbar\omega_1[eV]}{x}\left(\frac{m^*_p}{m}+ \frac{m^*_n}{m}\right)}\right)^3\phi(x),
 \end{equation}
 where
 $c_0=\frac{4\pi a^3}{3}\beta\frac{n_{s}}{H}$, with $n_{s}$ the surface density of metallic nanospheres, $H$ the semiconductor layer depth,
   $\phi(x)=\frac{x^2}{(x^2-1)^2+4x^2/x_1^2}\frac{1}{\sqrt{x-x_g}}$,
 $x=\omega/\omega_1$, $x_1=\tau^{(2)}_3\omega_1$, $x_g=E_g/(\hbar\omega_1)$, $\hbar\omega_1=2.72$  eV, $m_{n(p)}$ is the effective mass of conduction band and valence
 band
 carriers (for $Si$, $m^*_n=0.19(0.98)\;m$ and $m^*_p=0.16(0.52)\;m$, for light (heavy) carriers, band gap $E_g=1.14$ eV, $\varepsilon =12$), $m$ is the
 bare electron mass.

 The results are summarized in  Tab. 2 and in  Fig. 2,
for various radii of the nanospheres, and  reproduce well the experimental behavior reported in Ref.\onlinecite{wzr2}.
By $x_m$ we denote  frequencies corresponding to maximum value of the photo-current
 (i.e., to maximum of $I'/I$).

\vspace{2mm}

\begin{tabular}{|p{1.5cm}|p{2.5cm}|p{1cm}|p{4.5cm}|p{3cm}|p{1cm}|p{1.5cm}|}
\multicolumn{6}{l}{Tab. 2. Comparison with the experimental data\cite{wzr2} for Au nanospheres on Si layer \label{tab2}}\\
\hline
$a$ [nm] & $n_s$  [10$^8$/cm$^2$] & $x_m$ & $\omega_m= x_m\hbar\omega_1$ (theor) [eV] & $\hbar\omega_m$ (exp) [eV]&$\phi(x_m)$ &$\frac{I'}{I}(x_m)$\\
\hline
50  & 0.8& 0.772&2.09 &2.25 &0.84&1.55\\
\hline
40 & 1.6&0.951& 2.58& 2.48&3.00&1.9\\
\hline
25 & 6.6&  0.997& 2.71& 2.70&49.42&1.75\\
\hline
\end{tabular}\\

(the best coincidence with the experimental data  is attained at $\beta=3.5 \cdot  10^{-3} \frac{50^2}{(a[nm])^2}$)

\vspace{1mm}

In Fig. 2 an estimation of normalized photo-current, $I'/I$, with respect to wave-length is presented, for three sizes of metallic  nanospheres (Au)
deposited on photo-active Si layer,
with structure parameters as listed below (the proximity parameter $\beta=3.5 \cdot 10^{-3}\frac{50^2}{(a[nm])^2}$):

\begin{tabular}{|p{3cm}|p{4cm}|p{5cm}|p{3.5cm}|}
\multicolumn{4}{l}{Legend to Fig. 2}\\
\hline
 panel in Fig. 2&radius $a$ [nm]& concentration  $n_s$ [$10^8/cm^2$]  &  layer depth $H$ [$\mu$m]\\
 \hline
left&(A) 25, (B) 40, (C) 50 & (A)  6.6, (B) 1.6, (C) 0.8 &2\\
\hline
central &(A) 19, (B) 40, (C) 50  &(A)  6.6, (B) 1.6, (C) 0.8  &230 \\
\hline
right &(A) 25, (B) 40, (C) 50 &(A)  1.5, (B) 1.5, (C) 1.5  & 230\\
\hline
\end{tabular}\\

As it was indicated above, the relatively high value of  $\frac{q_m}{q}\sim 10^{5}\frac{\beta 40}{H[nm]}$ makes possible
 a significant  growth of efficiency of the photo-energy transfer to semiconductor, mediated by surface plasmons in nanoparticles deposited on the active layer,
by increasing $\beta$ or reducing $H$ (at constant $n_s$). However, because of the fact that
 an enhancement of $\beta$ easily induces the overdamped regime---cf. Eq. (\ref{tau}),
a more perspective would be thus
lowering of $H$, the layer depth  (cf. Fig.2 (left), where a significant growth of photo-current
with lowering of active layer depth $H$ illustrates the surface character of the effect).  The overall behavior of $I'/I(\omega)=1+q_m/q$  calculated according to the relation (\ref{fot}),
and depicted in the central panel in Fig. 2,
agrees  quite well with the experimental observations presented in Fig. 4 of Ref. \onlinecite{wzr2} (cf. inset in the central panel of Fig. 2),
both in position, height and shape of photo-current curves for distinct samples (the strongest enhancement is achieved for $a=40$ nm
at densities as indicated above, in the Legend to Fig. 2),
though $q_m/q$ is probably overestimated as the $q$ denominator would be
greater for doped real semiconductor structure which was not taken into account in the present calculus, similarly as surface
effects---all of these would change the $q$ denominator as well as its energy dependence, especially for longer wavelengths, where
the discrepancy between theoretical model and experimental data is noticeable.

 \section{Comments and conclusions}

 The presented analysis featuring semiclassical RPA-type approach to collective fluctuations in metallic nanosphere in 'jellium'
  model deals with two types of plasmons, surface and volume ones. Within this approximation
  the self-frequencies of surface plasmon modes are independent of the sphere radius (similarly as classical Mie frequency for dipole surface oscillations).
  There are, however, also surface modes induced by the volume modes and frequencies of
 these volume-induced surface oscillations  depend on the sphere radius, similarly as the self-frequencies of the volume plasmons
  (given by the dispersion relation  $\omega^2_{nl}= \omega_p^2(1+x_{nl}^2/(k_Ta^2))$, $x_{nl}$---nodes of the $l$th spherical Bessel function).
  The e-m response of the sphere consists
 of  both resonance types, the surface and the volume ones. It should be, however, emphasized that exciting of the volume
  modes is limited  by the nanoscale of the system
resulting in almost uniform e-m wave fields for resonant wavelength (dipole approximation regime).
The uniform over the sphere, dynamic
electric field excites the surface plasmons but not the radius-dependent volume modes.
Therefore one can conclude that the experimentally observed   significant dependence of resonant e-m frequencies on the radius of nanoparticles\cite{wzr2}
should be addressed
to more complicated phenomena than radius dependent volume modes.

The shift of the resonance frequency (in particular of Mie dipole type oscillations) for small clusters
 (up to $a\sim 2$ nm) was analyzed against various quantum effects in microscopic type approaches, mainly of
 TDLDA type\cite{brack,ekardt,weick} also within  semiclassical approaches\cite{kresin}. All
  these investigations indicate a major component of the experimentally observed red-shift of Mie frequency
  due to quantum spill-out effect (via reducing of density of electrons,  resulting in factor
  $\sqrt{1-\frac{\Delta N}{N_e}}$ for resonance frequency, where the spill-out
   volume  $\Delta N$, i.e., number of electrons outside the jellium edge, was the subject of various microscopic
   estimations\cite{brack,kresin,ekardt,weick}). Described in that manner red-shift of the resonance
    turns out, however, insufficient in comparison to experimental data\cite{kresin,ekardt,weick1}.
      It is  lower than observed in  the experiment for ultra-small clusters\cite{brack,kresin}.
      The contribution to red-shift, also important rather in small clusters, was obtained due to decay of
      plasmons for particle-hole pairs (Landau damping)\cite{ekardt1,weick1}, which improved
      fitting with the experiment. Additionally it was predicted an opposite blue-shift due to
       multi-plasmon anharmonic contribution\cite{gerch}.
       For larger cluster it was indicated that the dominating spill-out factor
       weakens as $\sim\frac{1}{a}$\cite{brack,kresin,ekardt,weick} and for
       considered in the present paper region of nanosphere dimensions (above 10 nm up to 60 nm)
       is negligible in comparison to predominant in this region of nanosphere radii radiation loss contribution
       (cf. also Fig 1).

Note. that within microscopic approach TDLDA
 more difficult it is to explain microscopically the observed width of resonance
 peak than its position\cite{brack}. The processes which
 would contribute are: spontaneous ionization, Landau damping (i.e.,
  interference of collective state with particle-hole pair with similar energy),
  evaporation of ion, ion vibrations (above jellium model). For larger $a$,  above
   10 nm, the dominating contribution to peak width\cite{kreib} is  caused by
   irradiation effects  which are stronger than scattering contribution
    $\sim  \frac{v_F}{\lambda_B} +\frac{Cv_F}{a}$
    (and C---constant of order 1 and $\lambda_B$---effective mean path,
    were estimated by many authors, cited in Ref. \onlinecite{brack},
    in particular including Landau damping type fragmentation of collective excitation).

The quantum spill-out, Landau damping and coupling to ion excitations (beyond jellium model), though important for small
clusters, are thus rather weak for large nanospheres and  contribute resonance shift for radius range
above 10 nm far lower than experimentally observed.     Another possibility for radius dependence of resonant frequencies is  connected
with the interaction of surface plasmons with other components of the system, which  leads to  damping of these oscillations.
A  shift of the resonance  for driven and damped oscillator depends on the attenuation rate, which scales with the nanosphere radius.
We have analyzed various channels of surface plasmon damping.  The most effective channel
for the surface plasmon damping is connected with the dipole-type near-field coupling of the
surface dipole plasmons with semiconductor substrate, on which  metallic nanospheres would be
deposited, e.g.,  in nanomodified diode-type systems. Due to nano scale of the spheres
for this coupling the momentum is not conserved, which results in a strong enhancement of the interband transition  probability
(because  all indirect electron transitions between valence and
conductivity bands in substratum have to be accounted for, provided  energy conservation alone).
It agrees with the  experimental data referred to a significant growth of the energy transfer from surface plasmons in metallic
nanoparticles to substrate semiconductor.

In order to include the damping of surface plasmons one can introduce a phenomenological damping factor  $\tau $ (attenuation time) to the oscillation
semiclassical RPA equation for electron local density fluctuations.
As the form of the resulting equation is of the damped oscillator type, thus  attenuation  causes
the red shift in the resonance frequency, $\omega'=\sqrt{\omega^2-\frac{1}{\tau^2}}$. For
driven and damped stationary oscillations, the red-shift of a resonance
takes place with maximal amplitude  at $\omega_m=\sqrt{\omega^2-\frac{2}{\tau^2}}$.
This red-shift is dependent on the sphere radius,
via  radius-dependence of $\tau$.

Energy transfer to semiconductor surroundings mediated by surface plasmons is so effective that it may  easily cause overdamped
 regime for plasmon oscillations. This channel is, however, reduced (typically by three orders in magnitude) by proximity constraints.
 Nevertheless, for nanospheres  deposited on the semiconductor surface even only small fraction  of the near-field
 channel ($\sim  \frac{h^2}{a^2} \sim 10^{-3} $, for $a\sim 50 $ nm,  $h$ is an effective range of  the
 near-field coupling) causes a strong  damping of plasmons.
 If to embed the nanospheres in semiconductor medium the
 plasmon system would fall in the overdamped regime ($\omega_1\tau \geq 1$).

 In the case of a small contact of the metallic nanospheres with
 the semiconductor substrate or in the case of an absence of semiconductor surroundings, a significant contribution  to
 plasmon attenuation is due to far-field radiation and electron scattering effects. The radiation contributions
 to $\frac{1}{\tau}$ scale with particle radius $a$ as $a^3$ (for both far- and near-field channels, though in the latter case
 the proximity constraints, included in $\beta$, would modify this dependence to the linear one), while for scattering contribution, $\frac{1}{\tau}\sim \frac{1}{a}$.
 Thus the total attenuation rate  $\frac{1}{\tau}\sim A a^3 + B a+ C \frac{1}{a}$ ($A, B, C$ constants).
  For relatively  big spheres  ($ a>10$ nm) the radiation channels prevail, while for smaller ones the
 scatterings would be also important\cite{stietz,scharte}.

The reported strengthening of photo-voltaic effects due to plasmonic concentrators
  (layer of metallic nanoparticles on active semiconductor
 surface with $n_s$ of order of $10^8$/cm$^2$), for instance:
up to  20-fold increase of photo-current in Si with nanoparticles Ag (40 nm [2-fold increase], 66 nm [8-fold], 108 nm [20-fold])\cite{wzr1},
indicate the significant role of near-field energy transfer growing with the sphere radius.
 Another observations confirm also the strengthening
role of plasmonic oscillations for emission and absorption phenomena in semiconductor diode systems, e.g.,
9-fold increase of emission from Si diode modified with nanoparticles  Ag of elliptical shape 120x60 nm, and resonant emission shift
after covering  nanoparticles  Ag  with 30 nm layer of ZnS\cite{wzr4,wzr5} and up to 14-fold increase of absorption with various metal nanoparticles:
 Ag (12 nm [3-fold], Au (10 nm [5-fold], Cu (10 nm [14-fold])\cite{wzr3}.
 Influence of dielectric coating is caused by  the surface self-oscillation sensitivity to
 dielectric surroundings (as for metallic nanosphere embedded in a dielectric medium), $\omega_{0l}=\omega_p
 \sqrt{\frac{l}{2l+1}\frac{1}{\varepsilon}}$, and for typical $\varepsilon \sim 10$ it gives strong decrease of resonant frequencies by factor $\sim 0.3$.
 The best correspondence with the experiment is attained for
 reported  strong dependence of extinction features with respect to nanoparticle size (located on the surface of  Si),
 the shift of the resonant peak   corresponding to the change of Au nanoparticles radius: 25--50 nm (stronger extremum for 40 nm)\cite{wzr2}
 and simultaneous enhancement of photo-current seem to be well described by our model.

Some experimental data indicate, however,  existence of  competitive mechanisms. For instance, for active medium TiO$_2$ the photo-current diminishes
in a wide spectral region (excluding the UV range) for coverings with nanoparticles Ag (3--6 nm), while the same coverings on optically active organic  medium
 (DSC---dye solar cell) lead to strong increase of the photo-current for 3 nm Ag, but to the decrease of photo-current for  6 nm Ag\cite{wzr4}.
 The competitive factors can be linked here with retardation of the carriers transfer, despite plasmonic strengthening, or with the
 destructive modification of a photo-sensitive substrate material, for too small nanoparticles  (more convenient are probably greater nanoparticles\cite{wzr1}).

  Summarizing, in the presented model nanosphere surface plasmons couple with substrate charges (band electrons in a substrate semiconductor)
  via photon-less
  short range e-m dipole interaction  with very quick timing (thus very effective)---as confirmed by time-resolved
   spectroscopy measurements\cite{wzr2}.
  The strong enhancement of the efficiency results from nanoscale-induced incommensurability leading to all indirect in momentum interband
    transitions, not allowed for interaction of band electrons with the original incident  planar wave photons as in an ordinary photo-effect.
     The type of dipole coupling
    is connected here with a specific e-m field gauge in the vicinity of the nanosphere within  the distance lower than  the wave-length (thus 'inside' the
    single photon), crucially distinct than for the planar wave (in the latter case only vector potential can be used,
    which is impossible in the former case)\cite{lan}.
The schematically described above scenario qualitatively fits with the  experimentally observed behavior
 and elucidates the timing of the particular steps of the energy-transfer-processes including mediating role of metallic nanosphere surface plasmons.
 The relevant time rates  can be estimated within
 standard quantum mechanical attitude of Fermi-golden-rule type.  Thus the presented above RPA plasmon description  supply the convenient
 and  simple tool for  further modeling and optimization of the metallically nanomodified solar cell structures, towards enhancement
  of their efficiency.

 \begin{acknowledgments}
 Supported by the Polish KBN Project No: N N202 260734 and the FNP Fellowship START (W. J.).

\end{acknowledgments}

\appendix

   \section{Analytical solution of plasmon equations for the nanosphere }

\label{app1}

Below we present a method of solution of Eqs (\ref{e20},   \ref{e21}).
For the solution of Eq. (\ref{e20}) we assume  in accordance with the rotational symmetry:
            $ \delta \tilde{\rho}_1( {\bm r,t})=n_e\left[f_1(r)+F({\bm r}, t)\right], \;for\; r<a$
(with an initial condition $  F({\bm r}, t)=0|_{t=0}$). After substituting it into  Eq. (\ref{e20}) we obtain:
\begin{equation}
\label{ea1}
\begin{array}{l}
\nabla^2 f_1(r)-k_T^2  f_1(r) =0,\\
\frac{\partial^2 F({\bm r},t)}{\partial t^2}=\frac{v_F^2}{3}\nabla^2 F({\bm r},t) -\omega_p^2  F({\bm r},t),\\
\end{array}
\end{equation}
The function $f_1(r)$ (nonsingular at $r=0$) has thus the form:
\begin{equation}
\label{alpha}
f_1(r)=\alpha \frac{e^{-k_Ta}}{k_Tr}\left(e^{-k_Tr}  -  e^{k_Tr}\right),
\end{equation}
where $\alpha$--const.,  $k_T=\sqrt{\frac{6\pi n_e e^2}{\epsilon_F}}=\sqrt{\frac{3\omega_p^2}{v_F^2}}\;$     ($k_T$--inverse  Thomas-Fermi radius),
$\omega_p=\sqrt{\frac{4\pi n_e e^2}{m}}\;$ (bulk plasmon frequency).
For  $  F({\bm r}, t)$ we assume   a single harmonics $  F({\bm r}, t)= F_{\omega}({\bm r}) sin(\omega t)$, as for a linear  differential equation and
suitably to the initial condition. Thus from Eq. (\ref{ea1}) we obtain:
\begin{equation}
\nabla^2  F_{\omega}({\bm r})+k^2 F_{\omega}({\bm r})=0,
\end{equation}
with $k^2=\frac{\omega^2-\omega_p^2}{v_F^2/3}$.
A solution nonsingular at $r=0$ has the form:
\begin{equation}
 F_{\omega}({\bm r})=Aj_l(kr)Y_{lm}(\Omega),
 \end{equation}
 where $A$--constant, $j_l(\xi)=\sqrt{\pi/(2\xi)}I_{l+1/2}(\xi)$--the spherical Bessel function [$I_{l}(\xi)$--the Bessel function of the first order],
 $Y_{lm}(\Omega)$--the spherical function ($ \Omega$--the spherical angle).
Owing to the quasiclassical boundary condition,   $ F({\bm r}, t)|_{r=a}=0$, one has to demand $j_l(ka)=0$, which leads to the discrete values of $k=k_{nl}=x_{nl}/a$,
(where $x_{nl},\;\;n=1,2,3...$, are nodes of $j_{l}$), and next to the discretization of self-frequencies $\omega$:
\begin{equation}
\omega_{nl}^2=\omega_{p}^2\left(1+\frac{x_{nl}^2}{k_T^2a^2}\right).
\end{equation}
              The general solution for $F({\bm r},t)$ has thus the form
\begin{equation}
\label{ea10}
F({\bm r},t)=\sum\limits_{l=0}^{\infty}\sum\limits_{m=-l}^{l}\sum\limits_{n=1}^{\infty}
A_{lmn}j_l(k_{nl} r)Y_{lm}(\Omega)sin(\omega_{nl}t).
\end{equation}

The neutrality condition, $\int\rho({\bm r},t)d^3r=N_e$, with
$\delta\rho_2({\bm r},t)=\sigma(\omega, t)\delta(a+\epsilon -r) +n_ef_2(r), (\;\epsilon\rightarrow 0)$, can be rewritten as follows:
$-\int\limits_{0}^{a} dr r^2f_1(r)= \int\limits_{a}^{\infty} dr r^2f_2(r)$, $\int\limits_{0}^{a}d^3rF({\bm r},t )=0$, $\int d\Omega \sigma(\Omega,t)=0$.
Taking into account also  the continuity condition on the spherical particle  surface, $1+f_1(a)=f_2(a)$, one can obtain:
$f_2(r)=\beta e^{-k_T(r-a)}/(k_Tr)$ and it is possible to fit $\alpha$ (cf. Eq. (\ref{alpha}))  and $\beta$ constants:
$\alpha=\frac{k_Ta +1}{2}$, $\beta=k_Ta -\frac{k_Ta+1}{2}\left(1-e^{-2k_Ta}\right)$.
    From the condition    $\int\limits_{0}^{a}d^3rF({\bm r},t )=0$  and Eq. (\ref{ea10}) it follows  that $A_{00n}=0$,
    (because of $\int d\Omega Y_{lm}(\omega)=4\pi \delta_{l0} \delta_{m0}$).
Therefore the internal-volume electron fluctuations in the sphere have the form:
\begin{equation}
\begin{array}{l}
\rho_1({\bm r}, t)=\Theta(a-r)\left[n_e+\delta\rho_1({\bm r},t)\right]=  \\
n_e\Theta(a-r)\left[1-\frac{k_Ta+1}{2}e^{-k_T(a-r)}\frac{1}{k_Tr}\left(1-e^{-2k_Tr}\right)+
\sum\limits_{l=1}^{\infty}\sum\limits_{m=-l}^{l}\sum\limits_{n=1}^{\infty}
A_{nlm}j_l(k_{nl} r)Y_{nl}(\Omega)sin(\omega_{nl}t)\right]. \\
\end{array}
\end{equation}

Now let us solve the Eq. (\ref{e21}) for surface electron fluctuations assuming
 $\delta \tilde{\rho}_2( {\bm r,t})=n_e f_2(r)+\sigma(\Omega,t)\delta(r+\epsilon-a), \;for\; r\geq a,\; ( r\rightarrow a+)$.
In order to remove the Dirac delta functions we integrate
both sides of the Eq. (\ref{e21}) with respect to the radius length  ($\int\limits_{0}^{\infty}r^2dr...$)
 and then we take a limit to the sphere surface, $\epsilon\rightarrow 0$. It results in the following equation:
 \begin{equation}
 \label{ea11}
\begin{array}{l}
a^2\frac{\partial^2\sigma(\Omega,t)}{\partial t^2}=\lim_{\epsilon,\epsilon'\rightarrow 0}\left\{-\frac{2}{3m}\int\limits_{0}^{\infty}
dr\frac{\partial }{\partial r} \left\{ \left[\frac{3\epsilon_F}{5} n_e+\epsilon_F\left(\sigma(\Omega,t) \delta(a+\epsilon'-r)
+n_ef_2(r)\right)\right]r^2\delta(a+\epsilon-r)\right\} \right.\\
 -\frac{2\epsilon_F}{3m} \sigma(\Omega,t)  \int\limits_{0}^{\infty} dr r^2  \delta(a+\epsilon-r) \frac{\partial }{\partial r} \delta(a+\epsilon'-r)
 -\frac{2\epsilon_F}{m}a^2 n_ef_2^{'}(a)\\
 -\frac{\omega_p^2}{4\pi}   \int\limits_{0}^{\infty} dr r^2   \delta(a+\epsilon-r) \frac{\partial }{\partial r}  \int\limits_{a}^{\infty} dr_1 r_1^2
  \int d\Omega_1 \frac{\sigma(\Omega_1,t) \delta(a+\epsilon'-r_1)  +n_ef_2(r_1)}{|{\bm r}-{\bm r}_1|}\\
   \left.-\frac{\omega_p^2}{4\pi}   \int\limits_{0}^{\infty} dr r^2   \delta(a+\epsilon-r) \frac{\partial }{\partial r}  \int\limits_{0}^{a} dr_1 r_1^2
  \int d\Omega_1 n_e\frac{F({\bm r}_1,t)+f_1(r_1)}{|{\bm r}-{\bm r}_1|}\right\}.\\
  \end{array}
  \end{equation}
  Note that the last term in the above equation describes a coupling between volume and surface plasmon excitations. The two first terms
  of the right-hand-side of the Eq. (\ref{ea11}) vanish in the limit $\epsilon,\epsilon'\rightarrow 0,\;( \epsilon<\epsilon')$ due to the delta function properties,
  and the third term gives the contribution: $\frac{\omega_p^2n_e}{k_T^3}\beta(1+k_Ta)$.
  The fourth and fifth terms contribute in the following manner:
 \begin{equation}
 \label{ea12}
 \begin{array}{l}
  -\frac{\omega_p^2a^2}{4\pi}\sum\limits_{l=1}^{\infty}\sum\limits_{m=-l}^{l}\frac{4\pi l}{2l+1}
  Y_{lm}(\Omega)\int d\Omega_1 \sigma(\Omega_1,t)  Y_{lm}^{*}(\Omega_1),\\
  -\frac{\omega_p^2\beta n_e}{k_T^3}(1+k_Ta)+\omega_p^2n_e  \sum\limits_{l=1}^{\infty}\sum\limits_{m=-l}^{l}\sum\limits_{n=1}^{\infty}
  A_{lmn}\frac{l+1}{2l+1}Y_{lm}(\Omega)
  \int dr_1 \frac{ r_1^{l=2}}{a^2} j_l(k_{nl }r_1)sin(\omega_{nl}t).\\
  \end{array}
  \end{equation}
 In the derivation of the fourth term  we have used (for $a<r_1$):
    \begin{equation}
                    \frac{\partial}{\partial a}\frac{1}{\sqrt{a^2+r_1^2-2ar_1cos\gamma}}
                    =  \frac{\partial}{\partial a}  \sum\limits_{l=0}^{\infty}\frac{a^l}{r_1^{l+1}}P_l(cos\gamma)= \sum\limits_{l=0}^{\infty}
                    \frac{la^{l-1}}{r_1^{l+1}}P_l(cos\gamma),
                    \end{equation}
where $P_l(cos\gamma)$ is the Legendre polynomial [$P_l(cos\gamma)=\frac{4\pi}{2l+1}\sum\limits_{m=-l}^{l}
Y_{lm}(\Omega)Y^{*}_{lm}(\Omega_1)$], $\gamma$ is an angle between vectors ${\bm a}=a\hat{\bm r}$ and ${\bm r}_1$,
 while in the derivation of the fifth term ($a>r_1$):
 \begin{equation}
                    \frac{\partial}{\partial a}\frac{1}{\sqrt{a^2+r_1^2-2ar_1cos\gamma}}
                    =  \frac{\partial}{\partial a}  \sum\limits_{l=0}^{\infty}\frac{r_1^l}{a^{l+1}}P_l(cos\gamma)=
                    -\sum\limits_{l=0}^{\infty}\sum\limits_{m=-l}^{l} 4\pi \frac{l+1}{2l+1} \frac{r_1^l}{a^{l+2}}
                    Y_{lm}(\Omega)Y^{*}_{lm}(\Omega_1).
                    \end{equation}
In the derivation of the last term in the Eq. (\ref{ea12})
we used the explicit forms of $F({\bm r},t)$ and $f_1(r)$, together with the neutrality condition,
$\int\limits_{0}^{a} dr r^2 f_1(r)=-\int\limits_{a}^{\infty} drr^2 f_2(r)=-\frac{\beta}{k_T^3}(1+k_Ta) $.
The above described procedure leads to the equation for the surface plasmons:
\begin{equation}
\label{ea15}
\begin{array}{l}
\frac{\partial^2 \sigma(\Omega,t)}{\partial t^2}=- \sum\limits_{l=0}^{\infty}\sum\limits_{m=-l}^{l}  \omega_{0l}^2
Y_{lm}(\Omega)\int d\Omega_1  \sigma(\Omega_1,t)Y^{*}_{lm}(\Omega_1)\\
+\omega_p^2n_e  \sum\limits_{l=0}^{\infty}\sum\limits_{m=-l}^{l} \sum\limits_{n=1}^{\infty} A_{lmn}
\frac{l+1}{2l+1} Y_{lm}(\Omega)  \int\limits_{0}^{a}dr_1 \frac{r_1^{l+2}}{a^{l+2}}j_l(k_{nl}r_1)sin(\omega_{nl}t),\\
\end{array}
\end{equation}
where $\omega_{0l}^2=\omega_p^2\frac{l}{2l+1}$.
Taking into account the spherical symmetry, one can assume the solution of the above equation in the form:
\begin{equation}
 \sigma(\Omega,t) = \sum\limits_{l=0}^{\infty}\sum\limits_{m=-l}^{l}q_{lm}(t)Y_{lm}(\Omega).
 \end{equation}
 After substituting it into the Eq. (\ref{ea15}) we find:
 \begin{equation}
 \begin{array}{l}
\frac{\partial^2 q_{00}(t)}{\partial t^2}=0,\;\; for\; l=0,\\
\frac{\partial^2 q_{lm}(t)}{\partial t^2}=-\omega_{0l}^2q_{lm}(t)+ \sum\limits_{n=1}^{\infty}\omega_p^2 n_e A_{lmn}
\frac{l+1}{2l+1}  \int\limits_{0}^{a}dr_1 \frac{r_1^{l+2}}{a^{l+2}}j_l(k_{nl}r_1)sin(\omega_{nl}t),\;for\; l\geq1,\\
\end{array}
\end{equation}
The solutions  of the above pair of equations have the form:
\begin{equation}
 \begin{array}{l}
q_{00}(t)=0, \;(acc.\; to\;  initial\;condition),\\
 q_{lm}(t) =\frac{B_{lm}}{a^2}sin(\omega_{0l}t)+ \sum\limits_{n=1}^{\infty} A_{lmn}
\frac{(l+1)\omega_p^2}{l\omega_p^2-(2l+1)\omega_{nl}^2} n_e \int\limits_{0}^{a}dr_1 \frac{r_1^{l+2}}{a^{l+2}}j_l(k_{nl}r_1)sin(\omega_{nl}t).\\
\end{array}
\end{equation}
And finally,
\begin{equation}
\begin{array}{l}
\sigma(\Omega,t)=\sum\limits_{l=1}^{\infty}\sum\limits_{m=-l}^{l} Y_{lm}(\Omega) \frac{B_{lm}}{a^2}sin(\omega_{0l}t)\\
 +\sum\limits_{l=1}^{\infty}\sum\limits_{m=-l}^{l}
\sum\limits_{n=1}^{\infty} A_{nlm}
\frac{(l+1)\omega_p^2}{l\omega_p^2-(2l+1)\omega_{nl}^2}Y_{lm}(\Omega) n_e \int\limits_{0}^{a}dr_1 \frac{r_1^{l+2}}{a^{l+2}}j_l(k_{nl}r_1)sin(\omega_{nl}t).\\
\end{array}
\end{equation}

Additionally let us comment on the equation for the case of  nanosphere embedded in the dielectric medium with $\varepsilon>1$---cf. Eq. (\ref{e26}).
The solutions of this equations are the same as presented above, however, with the
 frequencies $\omega_{0l}$ modified as follows:  $\omega_{0l}=\omega_p\sqrt{\frac{l}{2l+1}\frac{1}{\varepsilon}}$.

\section{Calculation of time rate for far-field  dipole-type radiation of nanosphere surface plasmons}
\label{app2}

In order to estimate the attenuation coefficient due to far-field radiation losses one can
consider  damping of nanosphere plasmons rapidly excited  by  switching off the uniform electric field,  $E(t)=E_0[1-\Theta(t)]$.
The corresponding oscillations of the local electron density can be  described by the equations:
\begin{equation}
\frac{\partial^2\delta\rho_1({\bm r},t)}{\partial t^2} +\frac{2}{\tau^{(1)}}\frac{\partial\delta\rho_1({\bm r},t)}{\partial t}
=\frac{v_F}{3}\triangle  \delta\rho_1({\bm r},t)-\omega_p^2 \delta\rho_1({\bm r},t),
\end{equation}
for $r<a$, and
\begin{equation}
\label{surface}
\begin{array}{l}
\frac{\partial^2\delta\rho_2({\bm r},t)}{\partial t^2}   +\frac{2}{\tau^{(2)}_2}\frac{\partial\delta\rho_2({\bm r},t)}{\partial t}
=-\frac{2\epsilon_F}{3m}\nabla\left[\frac{3}{5}n_e + \delta\rho_2({\bm r},t)\hat{r}\right]\delta(a+\epsilon-r)\\
-\left[ \frac{\omega_p^2}{4\pi}\hat{r}\nabla\int d^3r_1\frac{1}{|{\bm r}-{\bm r_1}|}\left(\Theta(a-r_1)\delta\rho_1({\bm r_1},t)
+\frac{1}{\varepsilon}\Theta(r_1-a) \delta\rho_2({\bm r_1},t)\right)+\frac{en_e}{m}E_r(t)\right]\delta(a+\epsilon-r),\\
\end{array}
\end{equation}
for $r=a$ ($\epsilon\rightarrow 0$). For $E$  not dependent on $r$, the driving force $E(t)$  enters to  the second equation only
and  leads to  the driven solution corresponding to the
dipole surface plasmon oscillations $\delta\rho_2 =Y_{10}(\Omega)q_{10}(t)$. Similarly, one can conclude that for nanospheres the visible light does not excite
 nanosphere volume plasmons as within  the dipole approximation  the incident wave electric field is uniform all over the sphere, unless the dipole
  approximation does not hold (i.e., when $a\sim \lambda$).

For $E(t)=E_0[1-\Theta(t)]$ (the rapid switching off  the constant electric field $E_0$) the
solution of Eq. (\ref{surface}) has the form:
\begin{equation}
\label{cosinus}
q_{10}(t)=\sqrt{\frac{4\pi}{3}}\frac{en_e}{m\omega_1^2}E_0\left\{\begin{array}{l}
1,\;\; for\;\; t<0,\\
\left[cos(\omega'_1 t)+\frac{sin(\omega'_1 t)}{\omega'_1\tau^{(2)}_2}\right]e^{-t/\tau^{(2)}_2},\;\; for\;\;\ t\geq 0,\\
\end{array}\right.
\end{equation}
where $\omega'_1=\sqrt{\omega_1^2-\left(\frac{1}{\tau_2^{(2)}}\right)^2}$ and $\omega_1=\omega_p\sqrt{\frac{1}{3\varepsilon}}$ is undamped dipole self-frequency.

It is easy to calculate the loss of the total energy of the system ${\cal{A}}={\cal{E}}(t=0)- {\cal{E}}(t=\infty)$, i.e., by taking into account both
kinetic and potential energy of electron system. Only potential interaction energy of oscillating
electrons contributes, and
$ {\cal{E}}(t)=const. +\frac{e^2}{2\varepsilon}a^3q_{10}^2(t)$, [the time dependent part of energy is caused by
interaction of  excited electrons,
$\frac{q^2_{10}(t)e^2}{2\varepsilon}
\int d^3r_1,d^3r_2\frac{Y_{10}(\Omega_1)\delta(a+\epsilon_1-r_1)     Y_{10}(\Omega_2)\delta(a+\epsilon_2-r_2)}{|{\bm r}_1  -  {\bm r}_2|}$,
with $\epsilon_1,\;    \epsilon_2 \rightarrow 0$, $\epsilon_1>\epsilon_2$]. For $q_{10}$ given Eq. (\ref{cosinus}) one can find,
\begin{equation}
\label{eqA}
{\cal{A}=\cal{E}}(t=0)- {\cal{E}}(t=\infty)= \frac{e^2}{2\varepsilon}a^3\frac{4\pi}{3}\left(\frac{en_eE_0}{m\omega_1^2}\right)^2,
\end{equation}
since $ {\cal{E}}(t)=const. +\frac{e^2}{2\varepsilon}\frac{4\pi}{3} a^3 \left(\frac{en_eE_0}{m\omega_1^2}\right)^2
\left(cos\omega'_1t +\frac{sin\omega'_1t}{\omega'_1\tau^{(2)}_2}\right)^2 e^{-2t/\tau^{(2)}_2}$.

On the other hand, assuming that damping of oscillations is caused by  far-field radiation,
one can calculate the energy loss $\cal{A}$ using the Poynting vector ${\bm \Pi}=
\frac{v}{4\pi}{\bm E}\times {\bm B}$, with $v=c/\sqrt{\varepsilon}$. The scalar potential of
the e-m wave emitted by the surface plasmon dipole oscillations: $ \rho({\bm r}, t)=e q_{10}(t) Y_{10}(\Omega)\delta (a+\epsilon -r)$
is of the retarded  form,
$\phi ({\bm R}, t)=\int \frac{\rho\left({\bm r}, t-\frac{|{\bm R}-{\bm r}|}{v}\right)}{\varepsilon |{\bm R}-{\bm r}|} d^3r$, and
for $R\gg a$,  $\phi ({\bm R}, t)=\frac{1}{\varepsilon Rv}{\hat{\bm n}}\cdot \frac{\partial {\bm D}\left(t-\frac{R}{v}\right)}{\partial t}$,
here $\hat{\bm n}=\frac{\bm R}{R}$ and the dipole moment  ${\bm D}(t-\frac{R}{v})=\int {\bm r} \rho({\bm r}, t-\frac{R}{v})d^3r$.
In our case of surface plasmon dipole oscillations
\begin{equation}
\label{dipol}
{\bm D}\left(t-\frac{R}{v}\right)=eq_{10}\left(t-\frac{R}{v}\right)\int{\bm r}Y_{10}(\Omega)\delta (a+\epsilon-r)d^3r
 =\left[0,0,eq_{10}\left(t-\frac{R}{v}\right)\sqrt{\frac{4\pi}{3}}a^3\right].
 \end{equation}

Similarly, for the retarded vector potential we find
 ${\bm A}({\bm R}, t)=\frac{1}{Rc} \frac{\partial {\bm D}\left(t-\frac{R}{v}\right)}{\partial t}$,
 [because of $\frac{\partial \rho}{\partial t}=-div {\bm j}({\bm r}, t)$, $\phi({\bm R}, t)=
 -\frac{1}{\varepsilon Rv}\int (\hat{\bm n} \cdot {\bm r}) div {\bm j}({\bm r}, t-\frac{R}{v})d^3r =
 \frac{1}{\varepsilon Rv} \hat{\bm n} \int {\bm j}({\bm r}, t-\frac{R}{v})d^3r$  for the sphere and due to
 $div ({\bm j} (\hat{\bm n}\cdot {\bm r}))= (\hat{\bm n}\cdot {\bm r})div {\bm j}+{\bm j}\cdot \hat{\bm n }$,
 which gives $\frac{\partial {\bm D}\left(t-\frac{R}{v}\right)}{\partial t}=\int{\bm j}\left({\bm r}, t-\frac{R}{v}\right)d^3r$].

 Hence, for far-field radiation of surface plasmon dipole oscillations
 we have
 \begin{equation}
 {\bm B}=rot {\bm A}=-\frac{\sqrt{\varepsilon}}{c^2R}\hat{\bm n}\times\frac{ \partial^2{\bm D}}{\partial t^2},
 \end{equation}
 and
 \begin{equation}
 {\bm E}=-\frac{1}{c}\frac{\partial {\bm A}}{\partial t} -\nabla \phi=\frac{1}{\sqrt{\varepsilon}}{\bm B}\times \hat{\bm n},
 \end{equation}
 which corresponds to the planar wave, and
        ${\bm \Pi}=\frac{\hat{\bm n}}{4 \pi}\frac{\left|\frac{\partial^2 {\bm D}}{\partial t^2}\right|^2sin^2\Theta}{\varepsilon v^3R^2}$, ($\Theta$
        is the  angle between ${\bm D}$ and ${\bm R}$).
        Next, taking into account that
$ \frac{d \cal{A}}{d t}=\oint {\bm \Pi}\cdot d{\bm s}$, one can find
${\cal{A}} = \int\limits_0^\infty \frac{d{\cal{A}}}{dt} dt
 = \frac{2}{3 \varepsilon v^3}\int\limits_0^\infty \left( \frac{\partial ^2D_z(t-R/v)}{\partial t^2} \right)^2 dt$.
 For $D_z=e\sqrt{4\pi/3} a^3q_{10}$ as in Eq. (\ref{dipol}), with $ q_{10}$  given by Eq. (\ref{cosinus}), one can find in this manner,
 \begin{equation}
 \begin{array}{c}
 {\cal{A}}=\frac{2}{3}\frac{e^2}{\varepsilon v^3}\frac{4\pi}{3}a^6 \left(\frac{en_eE_0}{m\omega_1^2}\right)^2 (\omega'_1)^4
 \left[1+\left(\frac{1}{\omega'_1\tau^{(2)}_2}\right)^2\right]^2\\
 \times\int\limits_0^{\infty}dt \left[1+\left(-1+\left(\frac{1}{\omega'_1\tau^{(2)}_2}\right)^2 \right)
 sin^2\omega'_1t-\frac{2sin\omega'_1tcos\omega'_1t}{\omega'_1\tau^{(2)}_2}\right]e^{-2t/\tau^{(2)}_2}.
 \end{array}
 \end{equation}
 The latter integral equals to $\tau^{(2)}_2/4$, which together with, $(\omega'_1 \tau^{(2)}_2)^2+1= (\omega_1 \tau^{(2)}_2)^2$,
 leads to the expression:
\begin{equation}
{\cal{A}}=\frac{e^2}{6\varepsilon v^3}\frac{4\pi}{3}a^6\left(\frac{en_eE_0}{m\omega_1^2}\right)^2 \omega_1^4\tau^{(2)}_2.
\end{equation}
 Via a comparison with Eq. (\ref{eqA}), we finally find,
\begin{equation}
\label{dof}
 \omega_1\tau^{(2)}_2=3 \left(\frac{\sqrt{3}c}{a\omega_p}\right)^3.
 \end{equation}

 \section{Calculation of  damping time rate due to near-field  interaction of surface plasmons with  semiconductor substrate}

\label{app3}

For the near-field regime ($\lambda >R> a,\;\;\lambda\gg a$)  the vector potential
has the same form as previously for far-field since only condition $a\gg R$ was  used for its derivation\cite{lan},
${\bm A}({\bm R},t)=\frac{1}{Rc}\frac{\partial {\bm D}\left(t-\frac{R}{v}\right)}{\partial t}$. In the near-field region the e-m field is not  of
planar wave type and both vector and scalar potentials are needed to describe it.
The scalar potential attains the form $\phi({\bm R}, t)= -div \frac{{\bm D}\left(t-\frac{R}{v}\right)}{\varepsilon R}$,
(due to the Lorentz gauge condition\cite{lan}, $div{\bm A}=-\frac{ \varepsilon \partial \phi}{c \partial t}$). The resulting Fourier components
of fields ${\bm B_{\omega}}$ and ${\bm E_{\omega}}$  (i.e. for monochromatic ${\bm D}={\bm D}_0 e^{-i\omega\left(t-\frac{R}{v}\right)})$
can be thus represented in this case as\cite{lan}:
\begin{equation}
{\bm B_{\omega}}=\frac{ik}{\sqrt{\varepsilon}}[{\bm D}_0\times \hat{\bm n}]\left(\frac{ik}{R} - \frac{1}{R^2}\right)
e^{ik R},
\end{equation}
and
\begin{equation}
 {\bm E_{\omega}}=\frac{1}{\varepsilon}\left\{{\bm D}_0\left(\frac{k^2}{R}+\frac{ik}{R^2}-\frac{1}{R^3}\right) +\hat{\bm n}(\hat{\bm n}\cdot
 {\bm D}_0)  \left(-\frac{k^2}{R}-\frac{3ik}{R^2}+\frac{3}{R^3}\right)\right\} e^{ik R},
 \end{equation}
where we use  the notation for the retarded argument, $i\omega\left(t-\frac{R}{c}\right)=i\omega t -i k R$.
For near-field region $kR\ll 1$ one can neglect terms with $\frac{1}{R}$ and    $\frac{1}{R^2}$. Assuming also that for
near-field   $e^{ikR}=1$ one can obtain thus
$ {\bm B_{\omega}}=0$ and $ {\bm E_{\omega}}  =\frac{1}{\varepsilon R^3}\left[3\hat{\bm n}\left(\hat{\bm n}\cdot {\bm D}_0\right)- {\bm D}_0
\right]$, which corresponds to dipole electric field.

The dipole type near-field potential can be written as follows:
\begin{equation}
\label{a3}
\varphi({\bm R},t)=\frac{1}{\varepsilon R^2} {\bm n}\cdot {\bm D_0}sin(\omega t+\alpha)= w^+e^{i\omega t} +w^-e^{-i\omega t},
\end{equation}
where $w^+=(w^-)^*=\frac{e}{\varepsilon R^2}\frac{1}{2i}e^{i\alpha}{\bm n}\cdot {\bm D_0}$;  one can confine Eq. (\ref{a3}) only to the $w^+$ term corresponding to
energy absorption in the semiconductor.
Then,  according to the Fermi golden rule, the transition probability per time unit between states
 $\Psi_{1{\bm k_1}}({\bm r},t)=(2\pi)^{-3/2}exp\left[i{\bm k_1}\cdot{\bm r}-iE_{1}({\bm k_1})t/\hbar\right],\;\;
\Psi_{2{\bm k_2}}({\bm r},t)=(2\pi)^{-3/2}exp\left[i{\bm k_2}\cdot{\bm r}-iE_{2}({\bm k_2})t/\hbar\right]$ (semiconductor electron states
from the valence and conduction bands, respectively), equals
\begin{equation}
w({\bm k_1}, {\bm k_2})=\frac{2\pi}{\hbar} |<{\bm k_1}|w^+|{\bm k_2}>|^2\delta(E_1({\bm k_1})  - E_2({\bm k_2}) +\hbar \omega),
\end{equation}
where $ <{\bm k_1}|w^+|{\bm k_2}>=\frac{1}{(2\pi)^3}\int \frac{e}{\varepsilon 2i}e^{i\alpha}{\bm n}\cdot {\bm D_0}
\frac{e^{-i({\bm k_1}-{\bm k_2})\cdot R}}{R^2} d^3 R$.
Taking $z$ axis along the vector ${\bm q}={\bm k_2}-{\bm k_1}$, then ${\bm q}\cdot {\bm R}=qRcos\Theta_1$,
${\bm n}\cdot {\bm D_0}=D_0(cos\Theta cos\Theta_1+sin\Theta sin\Theta_1cos\phi_1)$ ($\Theta$ is an angle between ${\bm D_0}$ and
${\bm q}$).
Hence, $ <{\bm k_1}|w^+|{\bm k_2}>=\frac{e}{(2\pi)^3 2i\varepsilon}e^{i\alpha}D_0 \int\limits_0^{\infty}dR\int\limits_0^{\pi}sin\Theta_1d\Theta_1
\int\limits_0^{2\pi}d\phi_1[cos\Theta cos\Theta_1+sin\Theta sin\Theta_1cos\phi_1]e^{iqRcos\Theta_1}=
\frac{eD_0}{(2\pi)^2\varepsilon}e^{i\alpha}\frac{cos\Theta}{q}$, [as $\int\limits_0^{\pi}cos\Theta_1 sin\Theta_1 d\Theta_1 e^{ixcos\Theta_1}=
-i \frac{d}{dx}2\frac{sinx}{x}$], and probability of transition
$w({\bm k_1}, {\bm k_2})= \frac{e^2D_0^2}{(2\pi)^3\hbar\varepsilon^2}\frac{cos^2\Theta}{q^2}
\delta(E_1({\bm k_1})- E_2({\bm k_2}) +\hbar\omega)$. In order to include all possible initial and final states in semiconductor, the summation
with respect to ${\bm k_1}$ and  ${\bm k_2}$ has to be performed (including filling factors $f({\bm k_1})\simeq 1$ and $f({\bm k_2})\simeq 0$,
as well as  absorption and emission of energy). In the result
we arrive with the total transition probability in semiconductor per time unit
 $\delta w \simeq \int\frac{d^3k_1}{4\pi^3} \int\frac{d^3k_2}{4\pi^3}w({\bm k_1}, {\bm k_2}) $ caused by dipole surface plasmon oscillations on the
 single nanosphere.

Let us emphasize that due to absence of the momentum conservation
for the near-field dipole coupling  in a vicinity of the nanosphere  all interband transitions contribute, not only direct ones as for the interaction with the
planar wave. It results in strong enhancement of the transition probability for the near-field coupling in comparison to photon (planar waves) attenuation rate in
a semiconductor in an ordinary photo-effect.

For the simplest model band structure,
$E_1({\bm k_1})- E_2({\bm k_2}) +\hbar\omega =x+y-\gamma$, where $ x=\frac{\hbar^2k_1^2}{2m_p^*}$, $ x=\frac{\hbar^2k_2^2}{2m_n^*}$ and
$\gamma=\hbar\omega -E_g$ ($E_g$ is the semiconductor band gap) the integration over wave vectors gives the formula for
the total probability of the transition
\begin{equation}
\label{5000}
\delta w =\frac{e^2D_0^2 \mu \sqrt{m_p^*m_n^*}}{3(4\pi^3)^2\hbar^5\varepsilon^2} (\hbar\omega_1 -E_g),
\end{equation}
where $\mu = \frac{ m_p^* m_n^*}{  m_p^* + m_n^*}$.

Assuming now that the dipole plasmon oscillations correspond to the damped oscillations which were excited by the rapid switching off
the uniform electric field (as  in the Appendix \ref{app2}), $E(t)=E_0(1-\Theta(t))$,  with the dipole-type solution for electron
distribution  given by Eq. (\ref{cosinus}), we have
\begin{equation}
\begin{array}{l}
{\bm D}(t)=[0,0, D_0e^{-t/\tau^{(2)}_3} cos(\omega'_1 t) Y_{10}(\Omega)\delta(a+\epsilon -r)]
\end{array}
\end{equation}
with
\begin{equation}
D_0=  \frac{e^2n_e}{m\omega_1^2}E_0 \frac{4\pi}{3}a^3;
\end{equation}
in comparison to the Eq. (\ref{cosinus}) we have neglected here the second term $\frac{sin(\omega'_1 t)}{\omega'_1\tau^{(2)}_3}$ for
 ${\tau^{(2)}_3} \omega'_1$  well greater than unity.

 One can now estimate the total energy transfer to the semiconductor (assuming that the dominant channel of the
 dissipation is the near-field interaction with semiconductor substrate and neglecting here the small  shift of $\omega'_1$
 due to dissipation)
   \begin{equation}
   \label{str}
   {\cal{A}}=\beta \int\limits_{0}^{\infty}\delta w \hbar\omega_1 dt= \beta \hbar \omega_1 \delta w \tau^{(2)}_3/2= \beta
\frac{\mu e^6n_e^2  E_0^2a^6\tau^{(2)}_3\sqrt{m_n^*m_p^*}\hbar\omega_1(\hbar\omega_1-E_g)}{6(3\pi^2)^2m^2\omega_1^4\hbar^5\varepsilon^2},
   \end{equation}
   where $\beta$ accounts for  the proximity constraints which reduce the near-field contact of the sphere with the
   semiconductor medium; for the case of nanospheres deposited on the semiconductor layer surface $\beta \sim \frac{h^2}{a^2} \sim
   10^{-3}$, for $a\sim 50$ nm ($h$ is an effective range of the near-field coupling),
    for the nanospheres entirely embedded in semiconductor surroundings $\beta$ would enhance significantly.
Comparing the value given by the formula (\ref{str}) with the energy loss given by Eq. (\ref{eqA}) one can find
\begin{equation}
\label{tau}
\frac{1}{\tau^{(2)}_3\omega_1}= \beta \frac{e^2  a^3\mu \sqrt{m_n^*m_p^*} (\hbar\omega_1 -E_g)}{36\pi^5 \hbar^4\varepsilon}.
\end{equation}
 For nanospheres of Au deposited on Si layer  we obtain:
 \begin{equation}
\frac{1}{\tau^{(2)}_3\omega_1}= 0.0059\beta\left(\frac{a}{[nm]}\right)^3 \frac{\mu}{m}
 \frac{\sqrt{m_n^*m_p^*}}{m},
\end{equation}
for light(heavy) carriers in Si,  $m_n=0.19(0.98)\;m$, $m_p=0.16(0.52)\;m$, and $E_g=1.14$ eV, $\varepsilon=12$, $\hbar\omega_1=2.72$ eV.

\newpage

\vspace{1 cm}

\begin{figure}[tb]
\centering
\scalebox{0.4}{\includegraphics{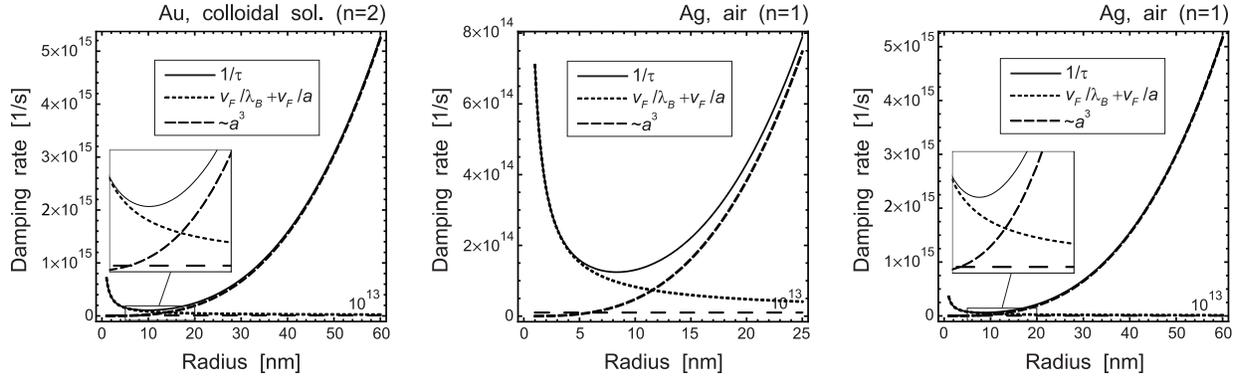}}
\caption{\label{fig1} Comparison of contributions to surface plasmon damping (upper curve) of
scattering term, $\sim \frac{v_F}{\lambda_B}+\frac{v_F}{a}$, and (far-field)  radiation losses-induced damping, $\sim a^3$,  Eq. (\ref{dof}), for large metallic
 nanospsheres (Au and Ag); for sphere radius  larger than
 10 nm  irradiation-induced-damping dominates (horizontal dashed line indicates $10^{13}$ level) }
\end{figure}

\begin{figure}[tb]
\centering
\scalebox{0.4}{\includegraphics{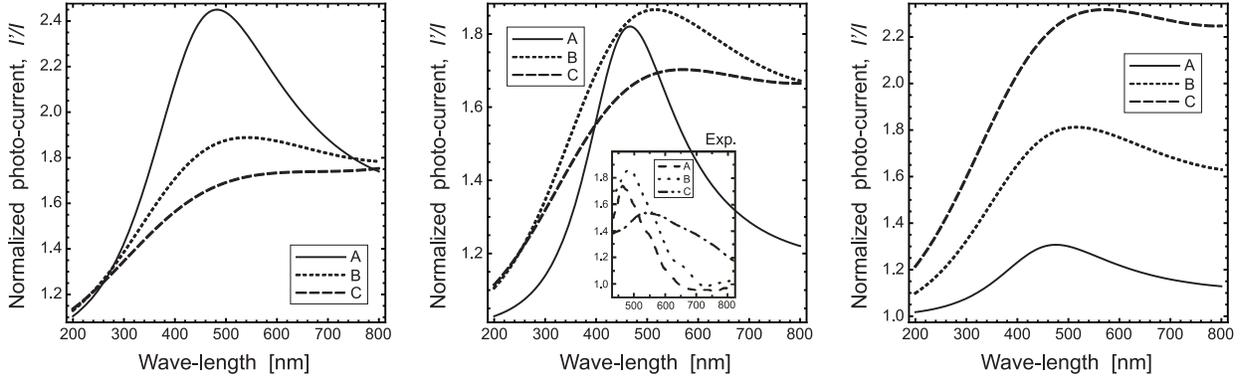}}
\caption{\label{fig2} Normalized photo-current $\frac{I'}{I}(\lambda)$ for various parameters\cite{wzr2}: $\beta=3.5 \cdot 10^{-3} \frac{50^2}{(a[nm])^2}$,
(left panel) $H=2\;\mu$m,  $a=25$ nm  (A), $40$ nm (B), $50$ nm (C), with
densities $n_s=6.6$ (A), $1.6$ (B), $0.8$ (C) $\times 10^8$/cm$^2$,
 (central panel) $H=230\;\mu$m,  $a=19$ nm  (A), $40$ nm (B), $50$ nm (C), with
densities $n_s=6.6$ (A), $1.6$ (B), $0.8$ (C) $\times 10^8$/cm$^2$, (right panel)
 $H=230\;\mu$m,  $a=25$ nm (A), $40$ nm (B), $50$ nm (C), with
densities $n_s=1.5$ (A), $1.5$ (B), $1.5$ (C) $\times 10^8$/cm$^2$, respectively;
coincidence with the experimental data\cite{wzr2} is achieved in the central panel, inset reproduces the experimental data\cite{wzr2}}
\end{figure}


\begin{thebibliography}{20}

\bibitem{plasmons}
	W. L. Barnes, A. Dereux, and T. W. Ebbesen, Nature {\bf 424}, 824 (2003)

\bibitem{maradudin}
A.V. Zayats, I. I. Smolyaninov, and A. A. Maradudin,  Phys. Rep. {\bf 408}, 131 (2005)

\bibitem{zastos}
	 S.A. Maier, {\it Plasmonics: Fundamentals and Applications} (Springer, Berlin 2007)


\bibitem{wzmocn1}
S. Pillai, et al.,  Appl. Phys. Lett. {\bf 88}, 161102 (2006)

\bibitem{wzmocn2}

M. Westphalen, U. Kreibig, J. Rostalski, H. L\"uth, and D. Meissner,
Sol. Energy Mater. Sol. Cells {\bf 61}, 97  (2000),
 M. Gratzel,  J. Photochem. Photobiol. C: Photochem. Rev. {\bf 4}, 145 (2003)


   \bibitem{wzr1}
H. R. Stuart and D. G. Hall, Appl. Phys. Lett. {\bf 73}, 3815 (1998);
H. R. Stuart and D. G. Hall, Phys. Rev. Lett. {\bf 80}, 5663 (1998);
  H. R. Stuart and D. G. Hall,  Appl. Phys. Lett. {\bf 69}, 2327 (1996)

\bibitem{wzr2}
	D. M. Schaadt, B. Feng, and E. T. Yu,  Appl. Phys. Lett. {\bf 86}, 063106 (2005)


\bibitem{wzr3}
	K. Okamoto, et al.,
  Nature Mat. {\bf 3}, 601 (2004); 	K. Okamoto, et al.,  Appl. Phys. Lett. {\bf 87}, 071102 (2005)

\bibitem{wzr4}
	C. Wen, K. Ishikawa, M. Kishima, K. Yamada, Sol. Cells {\bf 61}, 339  (2000)

\bibitem{wzr5}
P. Lalanne, J. P. Hugonin, Nature Phys. {\bf 2}, 551 (2006)

\bibitem{mie}
	G. Mie, Ann. Phys. {\bf 25}, 329 (1908)

\bibitem{6orders}
M. B. Mohamed, V. V. Volkov, S. Link, and M. A. El-Sayed, Chem. Phys. Lett. {\bf 317}, 517 (2000)


\bibitem{rodeffect}
G. T. Boyd, Z. H. Yu, and Y. R. Shen, Phys. Rev. B {\bf 33}, 7923 (1986)


\bibitem{hao}
E. Hao, R. C. Bayley, G. C. Schatz, J. T. Hupp, and S. Li, Nano Lett. {\bf 4}, 327 (2004)

\bibitem{burda}
C. Burda, X. Chen, R. Narayanan, M. A. El-Sayed,  Chem. Rev. {\bf 105}, 1025 (2005)





\bibitem{brack}
	M. Brack,  Rev. of Mod. Phys. {\bf 65}, 677 (1993)

\bibitem{kresin}  V. V. Kresin, Phys. Rep. {\bf 220}, 1 (1992)


\bibitem{ekardt}   W. Ekardt, Phys. Rev B {\bf 31}, 6360 (1985)

\bibitem{ekardt1} W. Ekardt, Phys. Rev. B {\bf 33}, 8803 (1986)

 \bibitem{weick}  G. Weick, R. A. Molina, D. Weinmann, and R. A. Jalabert, Phys. Rev. B {\bf 72}, 115410 (2005)

 \bibitem{weick1}  G. Weick, G. L. Ingold, R. A. Jalabert, and D. Weinmann, Phys. Rev. B {\bf 74},  165421 (2006)

\bibitem{serra}
L. Serra, F.  Garcias, M. Barranco, N. Barberan, and J. Navarro,
 Phys.   Rev. B {\bf 41}, 3434 (1990)

\bibitem{rubio}  A. Rubio and  L. Serra, Phys. Rev. B {\bf 48}, 18222 (1993)

\bibitem{gerch}  L. G. Gerchikov, C. Guet, and A. N. Ipatov, Phys. Rev. A {\bf 66}, 053202 (2002)



\bibitem{brack1}
 M. Brack,
   Phys. Rev. B {\bf 39}, 3533 (1989)



\bibitem{ekart}
W. Ekardt, Phys. Rev. Lett. {\bf 52}, 1925 (1984)






\bibitem{ksi}
C. F. Bohren, D. R. Huffman, {\it Absorption and Scattering of Light by Small Particles} (Wiley, New York, 1983)

\bibitem{kre}
 U. Kreibig,
M. Vollmer, {\it Optical Properties of Metal Clusters} (Springer, Berlin, 1995)


\bibitem{migdal}  A. B. Migdal, J. Phys. USSR {\bf 8}, 331 (1944)

\bibitem{steinw}  H. von Steinwedel and J. H. D. Jensen, Z. Naturforsh. A {\bf 5}, 413 (1950)

\bibitem{goldb} M. Goldhaber and E. Teller, Phys. Rev. {\bf 74}, 1046 (1948)


\bibitem{rpa} D. Pines, {\it Elementary Excitations in Solids} (ABP Perseus Books, Massachusetts, 1999)

\bibitem{pines}
D. Pines and D. Bohm, Phys. Rev. {\bf 85}, 338 (1952); D. Bohm and D. Pines, Phys. Rev. {\bf 92}, 609 (1953)

\bibitem{petrov} J. I. Petrov, {\it Physics of Small Particles} (Nauka, Moscow, 1984)


\bibitem{ag}
B. Lamprecht, A. Leitner, and F. R. Aussenegg, Appl. Phys. B: Lasers Opt. {\bf 64}, 269 (1997)




\bibitem{atwater}
M. L. Brongersma, J. W. Hartman, and H. A. Atwater, Phys. Rev. B {\bf 62}, R16356 (2000)





\bibitem{lan}
L. D.  Landau and E. M.  Lifshitz,  {\it Field Theory}  (Nauka, Moscow, 1973)



\bibitem{stietz}
F. Stietz {\it et al}, Phys. Rev. Lett. {\bf 84}, 5644 (2000)

\bibitem{scharte}
M. Scharte {\it et all.}, Appl. Phys. B: Laser Opt. {\bf 73}, 305 (2001)


\bibitem{kreib} U. Kreibig  and L. Genzel, Surf. Sci. {\bf c156}, 678 (1985)


 \bibitem{pol}
P. S. Kiriejew, {\it Physics of Semiconductors} (PWN, Warsaw, 1969)

\end{thebibliography}
\end{document}